%% file: resubmission.tex
\documentclass[12pt]{article}
\usepackage[utf8]{inputenc}
\usepackage[margin=1in]{geometry}
\usepackage{setspace}
\usepackage{amsmath}
\usepackage{amssymb}
\usepackage{hyperref}
\usepackage{graphicx}
\usepackage{caption}
\usepackage{natbib}
\usepackage{subcaption}
\usepackage{color}
\usepackage{float}

\input{notation}

\usepackage{amsthm}
\usepackage{dsfont}
\usepackage{mathrsfs}
\usepackage{lmodern}
\newtheorem{assumption}{Assumption}
\newtheorem{lemma}{Lemma}
\newtheorem{definition}{Definition}
\newtheorem{theorem}{Theorem}
\newtheorem{proposition}{Proposition}

\DeclareMathOperator*{\argmin}{arg\,min}
\providecommand{\keywords}[1]
{
  \small	
  \textbf{\textit{Keywords---}} #1
}
\DeclareMathOperator*{\argmax}{arg\,max}
\newcommand*{\QEDA}{\hfill\ensuremath{\square}}
\usepackage{appendix}
\title{Multi-agent Bayesian Learning with Best Response Dynamics: Convergence and Stability\thanks{This version: August, 2021.}}
\author{Manxi Wu, Saurabh Amin, and Asuman Ozdaglar\thanks{M. Wu is with the Institute for Data, Systems, and Society, S. Amin is with Laboratory for Information and Decision Systems, A. Ozdaglar is with Department of Electrical Engineering and Computer Science, Massachusetts Institute of Technology (MIT), Cambridge, MA, USA, \{manxiwu, amins, asuman\}@mit.edu}}
\date{}

\begin{document}

\maketitle
\begin{abstract}
We study learning dynamics induced by strategic agents who repeatedly play a game with an unknown payoff-relevant parameter. In this dynamics, a belief estimate of the parameter is repeatedly updated given players' strategies and realized payoffs using Bayes’s rule. Players adjust their strategies by accounting for best response strategies given the belief. We show that, with probability 1, beliefs and strategies converge to a fixed point, where the belief consistently estimates the payoff distribution for the strategy, and the strategy is an equilibrium corresponding to the belief. However, learning may not always identify the unknown parameter because the belief estimate relies on the game outcomes that are endogenously generated by players’ strategies. We obtain sufficient and necessary conditions, under which learning leads to a globally stable fixed point that is a complete information Nash equilibrium. We also provide sufficient conditions that guarantee local stability of fixed point beliefs and strategies.
\end{abstract}
\keywords{Learning in games, Bayesian learning, Stochastic dynamics, Stability analysis}
%
\input{introduction}
\input{model}

\input{fixed_point}
\input{convergence}

 \input{stability}



    
\input{MAP}
  \input{conclusion}

\section*{Acknowledgement}
We thank Vivek Borkar, Patrick Jaillet, Jason Marden, Shankar Sastry, John Tsitsiklis, Leeat Yariv, Muhammet Yildiz, Georges Zaccour for useful discussions.  We are grateful to speakers and participants at the 9th Workshop on Dynamic Games in Management Science at HEC Montréal (2017); 8th IFAC Workshop on Distributed Estimation and Control in Networked Systems (2019); 2nd Annual Conference on Learning for Dynamics and Control at Berkeley (2020).  This research was supported in part by Michael Hammer Fellowship, AFOSR project Building Attack Resilience into Complex Networks, and NSF CAREER award (CNS 1453126).  
\bibliographystyle{plainnat}
\bibliography{library.bib}

\appendix

\section{Supplementary Proofs for Section \ref{sec:main}}\label{apx:proof}

\input{proof_convergence}

 \input{proof_stability}

\end{document}

%% file: notation.tex
\newcommand{\E}{E}
\newcommand{\e}{e}

\renewcommand{\i}{i}

\newcommand{\deleq}{\stackrel{\Delta}{=}}

\newcommand{\loadep}{\delta^1}

\newcommand{\qi}{q_\i}

\newcommand{\rhothree}{\rho^3}
\newcommand{\qit}{\qi^\t}

\newcommand{\dist}{D}

\newcommand{\mi}{-\i}

\newcommand{\thetatil}{\tilde{\theta}}

\newcommand{\thetatilt}{\tilde{\theta}^{\t}}

\newcommand{\rhoone}{\rho^1}
\newcommand{\rhotwo}{\rho^2}

\newcommand{\Qmi}{Q_{-i}}

\newcommand{\Fi}{F_i}
\newcommand{\qtmi}{\qt_{\mi}}
\newcommand{\kt}{k_t}

\newcommand{\chat}{\c}

\newcommand{\Sequiv}{\S^{*}}
\newcommand{\qsran}{q^{*}}

\newcommand{\usi}{u^\s_{\i}}
\newcommand{\FP}{\Omega}

\renewcommand{\k}{t}

\newcommand{\neighinitheta}{N_{\thetaep}}
\newcommand{\neighiniload}{N_{\loadep}}
\newcommand{\neighinftheta}{N_{\thetaepfin}}
\newcommand{\neighinfload}{N_{\loadepfin}}

\newcommand{\qj}{q^j}

\newcommand{\Sbar}{\S\setminus [\thetabar]}
\newcommand{\q}{q}

\newcommand{\Q}{Q}
\newcommand{\Qi}{Q_\i}

\newcommand{\thetaep}{\epsilon^1}
\newcommand{\Uhat}{\hat{U}}
\newcommand{\loaddelta}{\delta}

\newcommand{\thetaepfin}{\bar{\epsilon}}
\newcommand{\loadepfin}{\epsilon_{\load}}

\newcommand{\shat}{\hat{\s}}

\newcommand{\qmi}{q_{-\i}}

\newcommand{\F}{F}

\newcommand{\sbar}{\bar{\s}}
\newcommand{\EQ}{\mathrm{EQ}}

\newcommand{\G}{G}

\newcommand{\epsilonhat}{\hat{\epsilon}}

\renewcommand{\loadepfin}{\bar{\delta}}
\renewcommand{\loaddelta}{\delta}

\renewcommand{\shat}{\hat{\s}}

\newcommand{\K}{K}

\newcommand{\Costhis}{Y^{\kt-1}}
\newcommand{\Loadhis}{Q^{\kt-1}}

\newcommand{\qbar}{\bar{\q}}

\newcommand{\phibar}{\phi}
\newcommand{\cbar}{\c}
\newcommand{\thetasran}{\theta^{*}}

\newcommand{\sran}{s^{*}}

\renewcommand{\(}{\left(}
\renewcommand{\)}{\right)}
\renewcommand{\c}{y}

\newcommand{\at}{a^\t}

\newcommand{\phis}{\phi^\s}
\renewcommand{\t}{k}

\renewcommand{\j}{j}

\newcommand{\pro}{\mathrm{Pr}}

\renewcommand{\a}{a}
\newcommand{\thetat}{\theta^\t}

\newcommand{\cj}{\cbar^j}

\newcommand{\Ai}{A_i}
\newcommand{\ai}{a_i}
\newcommand{\actiont}{a^\t}

\newcommand{\ait}{\at_\i}

\newcommand{\thetazero}{\theta^1}
\newcommand{\etaj}{\log\(\frac{\phibar^{\s}(\cj|\q^j)}{\phibar^{\sran}(\cj|\q^j)}\)}

\newcommand{\etabar}{\log\(\frac{\phibar^\s(\cbar|\qbar)}{\phibar^{\sran}(\cbar|\qbar)}\)}

\newcommand{\I}{I}

\newcommand{\qhat}{\hat{\q}}
\newcommand{\qhatt}{\hat{\q}^\t}

\newcommand{\n}{a}
\newcommand{\N}{A}

\newcommand{\epis}{\epsilon_{\i}^\s}

\newcommand{\T}{K}

\newcommand{\thetatone}{\theta^{\t+1}}

\newcommand{\Tstar}{\T^{*}}

\newcommand{\BR}{\mathrm{BR}}

\newcommand{\Chibarj}{\eta^j}
\newcommand{\s}{s}

\renewcommand{\S}{S}

\newcommand{\thetabar}{\bar{\theta}}

\newcommand{\ct}{\c^\t}

\newcommand{\qt}{\q^\t}

%% file: introduction.tex

\section{Introduction}
Strategic agents often need to engage in repeated interactions with each other while learning an unknown environment that impacts their payoffs. Such a situation arises in online market platforms, where buyers and sellers repeatedly make their transaction decisions while learning the latent market condition that governs the price distribution. The price distribution is updated based on the previous transactions and buyer reviews on platforms such as Amazon, eBay, and Airbnb (\cite{moe2004dynamic,acemoglu2017fast}). Another situation concerns with transportation networks, where travelers make route choice decisions on a day-to-day basis while also learning the underlying state of network that affects the travel time distribution. This travel time distribution is repeatedly updated based on the delay and flow information provided by navigation apps such as Google Maps or Apple Maps (\cite{zhu2010traffic, wu2021value, meigs2017learning}). In both situations, players' strategic decisions (purchases and sales on online platforms or route choices in transportation networks) influence the learning of the unknown environment (latent market condition or network state), which then impact the players' future decisions. Thus, the long-run outcome of strategic interactions among players is governed by the joint evolution of stage-wise decisions made by the players and learning of the unknown environment.  


In this article, we study learning dynamics that captures this joint evolution in a game-theoretic setting. In our model, strategic agents (players) repeatedly play a game with an unknown payoff-relevant parameter vector belonging to a finite set. A public information system (e.g. a market platform or navigation app) updates and broadcasts an estimate of the payoff parameter based on stage-wise game outcomes (i.e. strategies and randomly realized payoffs) to all players. The players' strategies in the next stage are given by a best response update rule based on the most current belief update. We consider three types of best response updates: \emph{Simultaneous best response}: each player best responds to their opponents' current strategy; \emph{Sequential best response}: players take turns to update their strategies in each stage; and \emph{Inertial best response}: the updated strategy is a linear combination of the current strategy and a best response strategy based on the updated belief. 

We focus on analyzing the long-run outcomes -- convergence and stability properties (both local and global) -- of the beliefs and strategies induced by the interplay of Bayesian updates and best response dynamics. We also identify conditions under which this learning dynamics converges to a complete information Nash equilibrium. Moreover, our technical results are useful to study other types of learning dynamics, such as learning under fast (resp. slow) strategy (resp. belief) updates, and learning under non-Bayesian estimates of the unknown parameter. 

Our model of learning dynamics leads to extension of results on learning in games with complete information to situations when long-run outcomes depend on learning of an unknown parameter. Past literature has addressed convergence analysis of discrete and continuous time best response dynamics (\cite{milgrom1990rationalizability, monderer1996potential, hofbauer2006best}), fictitious play (\cite{fudenberg1993learning, monderer1996fictitious}) and stochastic fictitious play (\cite{benaim1999mixed, hofbauer2002global}) in complete information environment. The stability properties of learning dynamics have also been studied in games with complete information (\cite{samuelson1992evolutionary, samuelson1994stochastic, sandholm2010local}). The key distinction between our learning dynamics and classical best response dynamics is that, in our model, players are imperfectly informed about the payoff-relevant parameter, and their strategy updates in each stage rely on the updated Bayesian belief. Our main contribution, as summarized next, is a new approach to study the convergence and stability properties of this learning dynamics.

\vspace{0.2cm}
\noindent\textbf{Convergence:} We prove that the beliefs and strategies in our learning dynamics converge to a fixed point with probability 1 (Theorem \ref{theorem:convergence}). Convergence is guaranteed under the condition that the best response dynamics converges in the game when the unknown parameter is constant instead of being repeatedly updated. This condition is satisfied in a variety of games, including potential games, zero-sum games and dominance solvable games. On the other hand, non-convergence can be easily demonstrated in games when this condition does not hold: if the strategies do not converge in a game with a constant belief, then natually they also fail to converge when the beliefs are repeatedly updated. 

Furthermore, at any fixed point, the belief consistently estimates the probability distribution of players' payoffs given the fixed point strategy and the fixed point strategy is an equilibrium of the game corresponding to the belief. When the true parameter is identifiable at equilibrium (i.e. no other parameter induces the same payoff distribution), the learning dynamics converges to a complete information equilibrium (Proposition \ref{prop:complete}). Otherwise, the fixed point belief may assign a non-zero probability to other parameters that lead to an incorrect payoff estimate for strategies that differ from the fixed point. Consequently, a fixed point strategy attained by the learning dynamics may not be a complete information equilibrium. 


The notion of fixed point in our learning dynamics is similar to the self-confirming equilibrium introduced in \cite{fudenberg1993self} for extensive-form games.\footnote{Similar concepts include conjectural equilibrium in \cite{hahn1978exercises} and subjective equilibrium in \cite{kalai1993subjective} and \cite{kalai1995subjective}.} At a self-confirming equilibrium, players maintain consistent beliefs of their opponents' strategies at information sets that are reached, but the beliefs of strategies can be incorrect at unreached information sets. Similarly, in our model, a fixed point can be different from a complete information equilibrium due to the incorrect estimates on the unobserved game outcomes formed by the beliefs (i.e., the estimated payoff distributions of strategies that differ from the fixed point may not be consistent). In general, these incorrect estimates may never be corrected by the learning dynamics because information of game outcomes is endogenously acquired based on the chosen strategies in each stage.\footnote{The phenomenon that endogenous information acquisition leads to incomplete learning is also central to multi-arm bandit problems \cite{rothschild1974two, easley1988controlling} and endogenous social learning \cite{duffie2009information, acemoglu2014dynamics, ali2018herding}.}


Notably, multiple models have been proposed as learning foundations for self-confirming equilibrium (\cite{fudenberg1993learning}, \cite{fudenberg1993steady}, \cite{kalai1993rational} and \cite{kalai1995subjective}). 
These models typically analyze how players maximize the present value of future payoffs in each stage of a repeated game with a fixed discount factor while updating the subjective beliefs of the opponents' strategies. Our learning model focuses on players' best response decisions that maximize their utilities based on the up-to-date knowledge of the payoff distribution and the opponents' strategies.

The proof of our convergence result uses techniques from statistical learning theory and analysis of best response dynamics in complete information games. Firstly, we obtain the convergence of Bayesian beliefs by applying the martingale convergence theorem. Secondly, thanks to the convergence of beliefs, we prove that the sequence of strategies induced by the learning dynamics with repeated belief updates converge to an auxiliary sequence of strategies that is constructed based on the convergent fixed point belief. Since the auxiliary strategy sequence converges to an equilibrium corresponding to the fixed point belief, we obtain that the original strategy sequence must also converge to the same fixed point equilibrium. Finally, using the strategy convergence result, we prove that the players' payoff distributions asymptotically approaches the identical and independent distribution generated using the fixed point strategy. This also allows us to show that the belief concentrates exponentially fast on the subset of parameters with the property that, at fixed point, each parameter in this set induces the same payoff distribution as the true parameter.


Our convergence result also contributes to the extensive literature on other types of learning dynamics: log-linear learning (\cite{blume1993statistical}, \cite{marden2012revisiting}, \cite{alos2010logit}), regret-based learning (\cite{hart2003regret}, \cite{foster2006regret}, \cite{marden2007regret}, \cite{daskalakis2011near}), payoff-based learning (\cite{cominetti2010payoff}, \cite{marden2009payoff}), replicator dynamics (\cite{beggs2005convergence}, \cite{hopkins2002two}), and learning in large anonymous games (\cite{kash2011multiagent, adlakha2013mean}). These dynamics typically prescribe the manner in which the players adjust their strategies based on the randomly realized payoffs in each stage. On the other hand, the strategy updates in our learning dynamics capture a rational behavioral adjustment of players in an imperfect information environment. 

\vspace{0.2cm}
\noindent\textbf{Stability:} 
We define a fixed point to be \emph{globally stable} if the learning dynamics starting from any initial state converges to that fixed point with probability 1. A fixed point is \emph{locally stable} if the states remain close to the fixed point with high probability when the learning dynamics starts with an initial state close to that fixed point. These stability notions apply to the coupled belief-strategy dynamics in a game theoretic setting.\footnote{\cite{frick2020stability} defined a similar stability notion for Bayesian beliefs of a utility-maximizing decision maker in a misspecified learning dynamics. In their problem, the unknown parameters can be ordered and information is endogenously acquired by the decision maker. We do not consider misspecification in our learning model. Our local stability notion is defined for the fixed point state (comprising of belief and equilibrium strategy) under the coupled updates of Bayesian beliefs and best response strategies in a game-theoretic setting.} 

We find that globally stable fixed points exist if and only if all fixed points have complete information of the unknown parameter (Proposition \ref{prop:global}). This condition is equivalent to the environment in which the true parameter is identifiable for any equilibrium strategy profile. In this case, all players eventually learn the true parameter and choose the complete information equilibrium. 

If the condition for global stability is not satisfied, then there exist multiple fixed points and convergent fixed point depends on the initial state. In this case, we need to analyze the local stability property of a fixed point, which entails studying conditions under which the states of learning dynamics after local perturbations remains close to the fixed point with a high probability. 

In Theorem \ref{theorem:stability}, we prove that a fixed point is locally stable if it satisfies three conditions: \emph{(a)} Fixed point strategy is \emph{locally upper-hemicontinuous} in the belief; \emph{(b)} Fixed point has a local neighborhood that is an \emph{invariant set} for the best response updates; \emph{(c)} Fixed point belief is \emph{locally consistent} in that it consistently estimates the payoff distribution in a local neighborhood of the fixed point strategy (instead of just at the fixed point). 

Previously, local stability of Nash equilibrium has been studied in games with complete information for both best response dynamics and evolutionary dynamics (\cite{smith1973logic, taylor1978evolutionary, samuelson1992evolutionary, matsui1992best, hofbauer2009stable, sandholm2010local}). We provide sufficient conditions that extend previous results to ensure stability of fixed point under local perturbations of the coupled belief-strategy dynamics. In particular, by using the martingale upcrossing inequality, we show that condition \emph{(c)} of local consistency ensures that the repeatedly updated beliefs remain in a small neighborhood of the fixed point belief with high probability. Additionally, condition \emph{(b)} extends the local invariance condition in complete information environment to further require that best response dynamics does not leave a local neighborhood of the fixed point under local perturbations of both the strategy and belief. Finally, condition \emph{(a)} of local upper-hemicontinuity ensures that the convergent strategy remains close to the fixed point as long as the belief is also close to the fixed point belief. Thus, local stability of a fixed point is guaranteed under conditions \emph{(a)} -- \emph{(c)}.

Next, we present our model and results in games with continuous strategy set: Section \ref{sec:basic_model} describes the learning model and Section \ref{sec:main} details the convergence and stability properties. In Section \ref{sec:variant}, we discuss the extensions of our main results to other types of learning dynamics such as two-timescale learning, learning in games with finite strategies, and learning with maximum a posteriori or least square estimates.

%% file: model.tex
\section{Model of Learning Dynamics in Continuous Games}\label{sec:basic_model}
Our learning dynamics is induced by strategic players in a finite set $\I$ who repeatedly play a game $\G$ for an infinite number of stages. The players' payoffs in game $\G$ depend on an \emph{unknown} parameter vector $\s$ belonging to a finite set $\S$. The true parameter is denoted $\sran \in \S$. Learning is mediated by a public information system (or an aggregator) that repeatedly updates and broadcasts a belief estimate $\theta=\(\theta(\s)\)_{\s \in \S} \in \Delta(\S)$ to all players, where $\theta(\s)$ denotes the estimated probability of parameter $\s$. 

In game $\G$, the strategy of each player $\i \in \I$ is a finite dimensional vector $\qi$ in a convex and continuous strategy set $\Qi$. The players' strategy profile is denoted $\q=\(\qi\)_{\i \in \I} \in \Q \deleq \prod_{\i \in \I} \Qi$. The payoff of each player is realized randomly according to a probability distribution. Specifically, the distribution of players' payoffs $\c=\(\c_i\)_{\i\in \I}$ for any strategy profile $\q \in \Q$ and any parameter $\s \in \S$ is represented by the probability density function $\phis(\c|\q)$. We assume that $\phis(\c|\q)$ is continuous in $\q$ for all $\s \in \S$. Without loss of generality, we write the player $\i$'s payoff $\c_{\i}$ for any $\s \in \S$ as the sum of an average payoff $\usi(\q)$ that is a continuous function of $\q$ and a noise term $\epis(\q)$ with zero mean: 
\begin{align}\label{eq:utility}
    \c_i=\usi(\q)+\epis(\q).
\end{align}
The noise terms $\(\epis(\q)\)_{\i \in \I}$ can be correlated across players. 

In game $\G$ with belief $\theta$, each player $\i$'s best response correspondence given their opponents' strategies $\qmi= \(\q^j\)_{j \in \I \setminus \{\i\}}$ is the set of strategies that maximize their expected utility, i.e. $\BR_i(\theta, \qmi)$ $\deleq \argmax_{\qi \in \Qi} \mathbb{E}_{\theta}\left[u_i^s(\qi, \qmi)\right] = \argmax_{\qi \in \Qi} \sum_{\s \in \S} \theta(\s)u_i^s(\qi, \qmi)$. Additionally, the set of equilibrium strategies for any belief $\theta$ is a non-empty set $\EQ(\theta)$.

Our learning model can be specified as a discrete-time stochastic dynamics, with state comprising of the belief estimate of unknown parameter and the players' strategies: In each stage $\t \in \mathbb{N}_{+}$, the information system broadcasts the current belief estimate $\thetat$; the players act according to a strategy profile $\qt=\(\qt_i\)_{\i \in \I}$; and the payoffs $\ct=\(\ct_i\)_{\i \in \I}$ are realized according to $\phi^s(\ct|\qt)$ when the parameter is $\s \in \S$. The state of learning dynamics in stage $\t$ is $\(\thetat, \qt\) \in \Delta(\S) \times \Q$. 

The initial belief $\thetazero$ in our learning dynamics does not exclude any possible parameter, i.e. $\thetazero(\s)>0$ for all $\s \in \S$, and the initial strategy $\q^1 \in \Q$ is feasible. The evolution of states $\(\thetat, \qt\)_{\t=1}^{\infty}$ is jointly governed by belief and strategy updates, which we introduce next.

\vspace{0.2cm}
\noindent\textbf{Belief update.} In our model, the belief is updated intermittently and infinitely. The stages at which the information system updates the belief can be deterministic or random, denoted by the subsequence $\(\kt\)_{\k=1}^{\infty}$. In update stage~$\t_{t+1}$, the previous belief estimate $\theta^{\kt}$ is updated using players' strategy profiles $\(\qt\)_{\t=\kt}^{\t_{t+1}-1}$ and realized payoffs $\(\ct\)_{\t=\kt}^{\t_{t+1}-1}$ between the stages $\t_{t}$ and $\t_{t+1}$ according to the Bayes' rule: 
\begin{align}
    \theta^{k_{t+1}}(\s)&= \frac{\theta^{\kt}(\s)\prod_{\t=\kt}^{\t_{t+1}-1}\phibar^\s(\ct|\qt)}{\sum_{s' \in \S} \theta^{\kt}(\s') \prod_{\t=\kt}^{\t_{t+1}-1}\phibar^{\s'}(\ct|\qt)}, \quad \forall \s \in \S. \tag{$\theta$-update} \label{eq:update_belief}
\end{align}

\vspace{0.2cm}
\noindent \textbf{Strategy update.} Players update their strategies in each stage based on the updated belief and the current strategies played by their opponents. Given any $\thetatone$ and any $\qtmi=\(\qt_j\)_{j \in \I \setminus \{\i\}}$, we generically denote the strategy update for each $\i \in \I$ as a set-valued function $\Fi\(\thetatone, \qtmi\): \Delta\(\S\) \times \Q_{-\i} \rightrightarrows \Qi$: 
\begin{align}\tag{$\q$-update}\label{eq:generic}
    \qi^{\t+1} \in \Fi\(\thetatone, \qtmi\), \quad \forall \i \in \I.
\end{align}
In particular, we consider the following three types of best response update rules for $\Fi$: 
\begin{enumerate}
    \item \emph{Simultaneous best response dynamics.} Each player chooses a strategy that is in the best response correspondence given their opponents' strategies and the updated belief: 
    \begin{align}\tag{Simultaneous-BR}\label{eq:sbr}
        \Fi(\thetatone, \qtmi)= \BR_i(\thetatone, \qtmi), \quad \forall \i \in \I.  
    \end{align}
    \item \emph{Sequential best response dynamics.} In each stage, exactly one player updates their strategy as the best response strategy given the new belief. Players sequentially updates their strategies: 
        \begin{align}\tag{Sequential-BR}\label{eq:br}
        \Fi(\thetatone, \qtmi)=\left\{\begin{array}{ll}  \BR_i(\thetatone, \qtmi), &\quad \text{if } \t ~ \text{mod} ~|\I| = \i,  \\
        \left\{\qi^{\t}\right\},& \quad \text{otherwise}.  
        \end{array}
        \right.
    \end{align}
    \item \emph{Linear best response dynamics.} Each player updates their strategy as a linear combination of their current strategy and a best response strategy given the updated belief: 
    \begin{align}\tag{Linear-BR}\label{eq:br_diminish}
    \Fi(\thetatone, \qtmi)&=(1-\alpha^k)\qit+ \alpha^{\t} \BR_i(\thetatone, \qt_{\mi}), \quad \forall \i \in \I, \quad \forall \t,  
\end{align}
where $\alpha^\t \in [0, 1]$ is the rate of strategy update in stage $\t$.
\end{enumerate}

 Next, we present few remarks about our learning dynamics: Firstly, players are strategic in that their strategy updates utilize a best response strategy that maximizes their expected utilities given the latest belief estimate of the unknown parameter and the strategies played by their opponents. If all players know the true parameter $\sran$, then the three strategy updates reduce to the classical best response dynamics in the corresponding game with complete information. 
 
 Secondly, the three types of strategy updates differ in the timing and the extent at which best response is incorporated in the updated strategy: All players update their strategies in every stage in \eqref{eq:sbr} and \eqref{eq:br_diminish}, while only one player updates strategy in \eqref{eq:br}. While players entirely adopt the new best response strategy in updates of \eqref{eq:sbr} and \eqref{eq:br}, in \eqref{eq:br_diminish} each player weighs their best response strategy according to the strategy update rate. 

Thirdly, the belief updates can occur less frequently than the strategy updates since the subsequence of belief update stages satisfy $\t_{t+1}- \kt \geq 1$. Here, we assume that $\t_{t+1}- \kt$ is finite with probability (w.p.) 1; i.e., both belief and strategy updates follow the same timescale. In Sec. \ref{sec:variant}, we extend our analysis to the case when belief updates occur at a slower timescale in comparison to strategy updates, i.e. $\lim_{t\to \infty} \t_{t+1}- \kt = \infty$. 

Fourthly, our learning dynamics considers games with continuous utility functions and strategy sets. In Sec. \ref{sec:variant}, we show that our convergence and stability results also apply to learning the unknown parameter in games with finite action (pure strategy) set, where players choose mixed strategies. As a special case, for games with finite strategies, the linear best response dynamics \eqref{eq:br_diminish} with update rates $\alpha^{\t}=\frac{1}{\t}$ for all $\t$ is equivalent to fictitious play with repeatedly updated belief estimates.  Finally, in Sec. \ref{sec:variant}, we also argue that our results hold under other types of parameter estimates such as the maximum a posteriori (MAP) estimate, and the ordinary least square (OLS) estimate.

%% file: convergence.tex
\section{Main Results}\label{sec:main}
\subsection{Convergence}\label{subsec:convergence}
Before introducing our convergence result, we introduce two necessary definitions.
\begin{definition}[Kullback–Leibler (KL)-divergence]\label{def:KL}
For a strategy profile $\q\in\Q$, the KL divergence between the payoff distribution with parameters $\s$  and $\sran\in\S$ is defined as:
\begin{align*}
     D_{KL} \left(\phibar^{\sran}(\cbar|\q)||\phibar^{\s}(\cbar|\q)\right) \deleq \left\{
     \begin{array}{ll}
     \int_{\chat} \phibar^{\sran}(\chat|\q) \log\left(\frac{\phibar^{\sran}(\chat|\q)}{\phibar^{\s}(\chat|\q)}\right) d\chat, & \quad \text{if $\phibar^{\sran}(\chat|\q) \ll \phibar^{\s}(\chat|\q)$}, \\
     \infty & \quad \text{otherwise.}
     \end{array}
     \right.
\end{align*}
\end{definition}
Here $\phibar^{\sran}(\chat|\q) \ll \phibar^{\s}(\chat|\q)$ means that the distribution $\phibar^{\sran}(\chat|\q)$ is absolutely continuous with respect to $\phibar^{\s}(\chat|\q)$, i.e. $\phibar^{\s}(\chat|\q)=0$ implies $\phibar^{\sran}(\chat|\q)=0$ w.p. 1.

\begin{definition}[Payoff-equivalent parameters]\label{def:payoff_equivalence}
A parameter $\s\in\S$ is payoff-equivalent to the true parameter $\sran$ for a strategy $\q\in\Q$ if $D_{KL} \left(\phibar^{\sran}(\cbar|\q)||\phibar^{\s}(\cbar|\q)\right)=0$. For a given strategy profile $\q\in Q$, the set of parameters that are payoff-equivalent to $\sran$ is defined as:  
\begin{align*}
    \Sequiv(\q)\deleq \{\S|D_{KL} \left(\phibar^{\sran}(\cbar|\q)||\phibar^{\s}(\cbar|\q)\right)=0\}.
\end{align*}
\end{definition}

The KL-divergence between any two distributions is non-negative, and is equal to zero if and only if the two distributions are identical. For a given strategy profile $\q$, if a parameter $\s$ is in the payoff-equivalent parameter set $\Sequiv(\q)$, then the payoff distribution is identical for parameters $\s$ and $\sran$, i.e. $\phibar^{\sran}(\chat|\q)=\phibar^{\s}(\chat|\q)$ for all $\chat$. In this case, realized payoffs cannot be used by the information system to distinguish $\s$ and $\sran$ in the belief update \eqref{eq:update_belief} (since the belief ratio $\frac{\thetat(\s)}{\thetat(\sran)}$ remains unchanged w.p. 1). Also note that the set $\Sequiv(\q)$ can vary with strategy profile $\q$, and hence a payoff-equivalent parameter for a given strategy profile may not be payoff-equivalent for another strategy profile.

We need the following assumption on the strategy updates.


\begin{assumption}\label{asu}
For any initial strategy $\q^1$, the sequence of strategies induced by \eqref{eq:generic} under any constant belief $\thetat=\theta \in \Delta\(\S\)$ for all $\t$ converges to an equilibrium strategy profile in $\EQ(\theta)$. 
\end{assumption}


This assumption requires that the strategy updates converge to an equilibrium strategy when the belief is held constant (instead of being repeatedly updated). Without this assumption, strategies may fail to converge even in games with complete information (\cite{shapley1964some}). Thus, Assumption \ref{asu} is a basic requirement to guarantee the convergence of states in our learning dynamics.

Assumption \ref{asu} is satisfied by the best response dynamics \eqref{eq:sbr}, \eqref{eq:br} and \eqref{eq:br_diminish} in a variety of games with complete information, including potential games, zero sum games, and dominance solvable games (\cite{milgrom1990rationalizability, monderer1996potential, hofbauer2006best}, \cite{fudenberg1993learning, monderer1996fictitious}). Under Assumption \ref{asu}, the sequence of states (beliefs and strategies) induced by our stochastic learning dynamics converges to a fixed point. 
\begin{theorem}\label{theorem:convergence}
For any initial state~$(\theta^1, \q^1)\in\Delta(S)\times \Q$, under Assumption \ref{asu}, the sequence of states $(\thetat, \qt)_{\t=1}^{\infty}$ induced by \eqref{eq:update_belief} and \eqref{eq:generic} converges to a fixed point $(\thetabar, \qbar)$ w.p. 1, and $\(\thetabar, \qbar\)$ satisfies: 
\begin{subequations}\label{eq:fixed_point_def}
\begin{align}
    [\thetabar] &\subseteq \Sequiv(\qbar), \label{eq:exclude_distinguished}\\
    \qbar&\in \EQ(\thetabar), \label{subeq:eq_fixed}
\end{align}
\end{subequations}
where $[\thetabar] \deleq \{\S|\thetabar(\s)>0\}$, and $\EQ(\thetabar)$ is the set of equilibrium strategies corresponding to belief $\thetabar$. 

Moreover, for any $\s \in \S\setminus \Sequiv(\qbar)$, if $\phibar^{\sran}(\chat|\qbar) \ll \phibar^{\s}(\chat|\qbar)$, then $\thetat(\s)$ converges to 0 exponentially fast: 
\begin{align}
    \lim_{\t \to \infty} \frac{1}{\t} \log(\thetat(\s))=-D_{KL}(\phibar^{\sran}(\cbar|\qbar)||\phibar^{\s}(\cbar|\qbar)), \quad  w.p.~1.\label{eq:rate}
\end{align}
Otherwise, there exists a positive integer $\Tstar<\infty$ such that $\thetat(\s)=0$ for all $\t>\Tstar$ w.p. 1. 
\end{theorem}


From Theorem \ref{theorem:convergence}, the following properties must hold at a fixed point $\(\thetabar, \qbar\)$:  
\begin{itemize}
    \item[(1)] Belief $\thetabar$ identifies the true parameter $\sran$ in the payoff-equivalent set $\Sequiv(\qbar)$ corresponding to fixed point strategy $\qbar$. Therefore, the belief forms a consistent estimate of the payoff distribution at the fixed point. To see this, let us denote the estimated payoff distribution as $\mu(\chat|\thetabar, \qbar)$. Then, 
\begin{align}\label{eq:marginal}
    \mu(\chat|\thetabar, \qbar)\deleq\sum_{\s \in \S} \thetabar(\s) \phibar^\s(\chat|\qbar)\stackrel{\eqref{eq:exclude_distinguished}}{=}\sum_{\s \in \Sequiv(\qbar)}\thetabar(\s) \phibar^{\s}(\cbar|\qbar)=\sum_{\s \in \Sequiv(\qbar)} \thetabar(\s) \phibar^{\sran}(\cbar|\qbar)= \phibar^{\sran}(\cbar|\qbar).
 \end{align}
 \item[(2)] Players have no incentive to deviate from fixed point strategy profile $\qbar$ because it is an equilibrium of the game $\G$ with fixed point belief $\thetabar$.
\end{itemize}

 We prove Theorem \ref{theorem:convergence} in three steps: Firstly, we prove that the sequence of beliefs $\(\thetat\)_{\t=1}^{\infty}$ converges to a fixed point belief $\thetabar \in \Delta\(\S\)$ w.p. 1 by applying the martingale convergence theorem (Lemma \ref{lemma:theta}). Secondly, we show that under Assumption \ref{asu}, the strategies $\(\qt\)_{\t=1}^{\infty}$ in our learning dynamics with belief updates also converge. This convergent strategy is an equilibrium corresponding to the belief $\thetabar$ (Lemma \ref{lemma:q}). Finally, we prove that the belief of any $\s \in \S$ that is not payoff-equivalent to $\sran$ given $\qbar$ must converge to $0$ with rate of convergence governed by \eqref{eq:rate} (Lemma \ref{lemma:consistency}). Hence, we can conclude that beliefs and strategies induced by the learning dynamics converge to a fixed point $\(\thetabar, \qbar\)$ that satisfies \eqref{eq:fixed_point_def} w.p. 1. The formal proofs of Lemmas \ref{lemma:theta} -- \ref{lemma:consistency} are in Appendix \ref{apx:proof}.

\begin{lemma}\label{lemma:theta}
$\lim_{\t \to \infty} \thetat=\thetabar$ w.p. 1, where $\thetabar  \in\Delta(S)$.
\end{lemma}

To prove this property, we note that the subsequences of the belief ratios $\(\frac{\theta^{\kt}(\s)}{\theta^{\kt}(\sran)}\)_{\k=1}^{\infty}$ is a martingale for all $\s \in \S$, and $\(\theta^{\kt}(\sran)\)_{t=1}^{\infty}$ is a sub-martingale. From the martingale convergence theorem, $\(\frac{\theta^{\kt}(\s)}{\theta^{\kt}(\sran)}\)_{\k=1}^{\infty}$ and $\(\theta^{\kt}(\sran)\)_{t=1}^{\infty}$ converge w.p. 1. Thus, the belief subsequence $\(\theta^{\kt}\)_{t=1}^{\infty}$ converges to a fixed point belief $\thetabar$ w.p. 1. Since $\thetat=\theta^{\kt}$ for any $\t = \kt, \dots, k_{t+1}-1$, the sequence $\(\thetat\)_{\t=1}^{\infty}$ must also converge to $\thetabar$.

\begin{lemma}\label{lemma:q}
Under Assumption \ref{asu}, $\lim_{\t \to \infty} \qt = \qbar$ w.p. 1, where $\qbar$ satisfies \eqref{subeq:eq_fixed}.
\end{lemma}

In the proof of Lemma \ref{lemma:q}, for each stage $\K=1, 2, \dots$, we construct an auxiliary strategy sequence $\(\qhatt\)_{\t=1}^{\infty}$ such that the strategies in this sequence are identical to that in the original sequence up to a certain stage $\K$ (i.e. $\qhatt=\qt$ for all $\t=1, \dots, \K$), and the remaining strategies $\(\qhatt\)_{\t=\K+1}^{\infty}$ are induced by the best response update with the fixed point belief $\thetabar$ (instead of the repeatedly updated belief sequence $\(\thetat\)_{\t=\K+1}^{\infty}$). Under Assumption \ref{asu}, the auxiliary strategy sequence must converge to an equilibrium $\qbar \in \EQ(\thetabar)$. Recall from Lemma \ref{lemma:theta}, the beliefs converge to $\thetabar$. Moreover, since the expected utility function $\mathbb{E}_{\theta}\left[u_i^s(\q)\right]$ of each player $\i \in \I$ is continuous in $\theta$ and $\q$, we know from the Berge's maximum theorem that the best response correspondence $\BR_i\(\theta, \q\)$ is upper hemicontinuous in $\theta$ and $\q$ (Lemma \ref{lemma:uh} in Appendix \ref{apx:proof}). Thus, we can prove that as $\K \to \infty$, the distance between the auxiliary strategy sequence and the original strategy sequence converges to zero, which implies that the original strategy sequence $\(\qt\)_{\t=1}^{\infty}$ also converges to $\qbar \in \EQ(\thetabar)$ (i.e. $\qbar$ satisfies \eqref{subeq:eq_fixed}).



\begin{lemma}\label{lemma:consistency}
Any fix point $\(\thetabar, \qbar\)$ satisfies \eqref{eq:exclude_distinguished}. Furthermore, for any $\s \in \S\setminus \Sequiv(\qbar)$, if $\phibar^{\sran}(\chat|\qbar) \ll \phibar^{\s}(\chat|\qbar)$, then $\thetat(\s)$ satisfies \eqref{eq:rate}.
Otherwise, there exists a finite positive integer $\Tstar$ such that $\thetat(\s)=0$ for all $\t>\Tstar$ w.p. 1. 
\end{lemma}

Lemma \ref{lemma:consistency} is based on Lemmas \ref{lemma:theta} and \ref{lemma:q}. Although the data of the realized payoffs $\(\ct\)_{\t=1}^{\infty}$ is not independently and identically distributed (i.i.d.) due to players' strategy updates, we can show that since $\qt$ converge to $\qbar$ (Lemma \ref{lemma:q}), the distribution of $\ct$ converges to an i.i.d. process with $\phi^{\s}(\ct|\qbar)$, which is the payoff distribution given the fixed point strategy $\qbar$, for each parameter $\s$ as $\t\to \infty$.

Finally, if the payoff distribution for the true parameter is absolutely continuous with respect to any non-payoff-equivalent parameter $\s \in \S\setminus \Sequiv(\qbar)$, then we show that the log-likelihood ratio $\log\(\frac{\thetat(\s)}{\thetat(\sran)}\)$ converges to $-\infty$ with an exponential rate given by the (non-zero) KL-divergence between the distributions of realized payoff under parameters $\s$ and $\sran$. Thus, the belief $\thetat(\s)$ converges to zero exponentially fast as in \eqref{eq:rate}. On the other hand, if $\phi^s(\c|\qbar)$ is not absolutely continuous with the true distribution $\phi^{\sran}\(\c|\qbar\)$, then we can find a small neighborhood of $\qbar$ such that, with positive probability,  the realized payoff $\c$ satisfies $\phi^{\s}(\c|\q)=0$ but $\phi^{\s*}(\c|\q)>0$ for $\q$ in this neighborhood. In this case, the belief update \eqref{eq:update_belief} will assign probability 0 to the parameter $\s$. From the Borel–Cantelli lemma, the probability that $\thetat(\s)$ remains positive infinitely often is zero. Hence, there must exist a finite stage $\Tstar$, after which $\thetat(\s)$ remains to be zero with probability 1.

\vspace{0.2cm}
\noindent\textbf{Complete information fixed points.} From \eqref{eq:fixed_point_def}, we define the set of fixed points $\FP$ as follows: 
\begin{align}\label{eq:FP}
    \FP \deleq \left\{\(\thetabar, \qbar\)\left\vert [\theta] \subseteq \Sequiv\(\qbar\), ~ \qbar \in \EQ(\thetabar)\right.\right\}.
\end{align}

We denote the belief vector $\thetasran$ with $\thetasran(\sran)=1$ as the \emph{complete information belief}, and any strategy $\qsran\in \EQ(\thetasran)$ as a \emph{complete information equilibrium}. Since $[\thetasran]=\{\sran\} \subseteq \Sequiv(\qsran)$, the state $\(\thetasran, \qsran\)$ is always a fixed point (i.e. $\(\thetasran, \qsran\) \in \FP$), and has the property that all players have complete information of the true parameter $\sran$ and choose a complete information equilibrium. Therefore, we refer to $\(\thetasran, \qsran\)$ as a \emph{complete information fixed point}.

Additionally, the set $\FP$ may contain other fixed points $\(\thetabar, \qbar\)$ that are not equivalent to the complete information environment, i.e. $\thetabar \neq \thetasran$. Such belief $\thetabar$ must assign positive probability to at least one parameter $\s \neq \sran$. The property \eqref{eq:exclude_distinguished} ensures that $\s$ is payoff-equivalent to $\sran$ given the fixed point strategy profile $\qbar$, and hence the average payoff function in \eqref{eq:utility} satisfies $\usi(\qbar)=u^{\sran}_i(\qbar)$ for all $\i \in \I$. However, for other strategies $\q \neq \qbar$, the value of $\usi(\q)$ may be different from $u^{\sran}_i(\q)$ for one or more players $\i \in \I$. That is, belief $\thetabar$ consistently estimates the payoff at a fixed point strategy $\qbar$ but not necessarily at all $\q \in \Q$. Consequently, if one or more players had access to complete information of the true parameter $\sran$, they may have an incentive to deviate from the fixed point strategy; such a fixed point strategy profile $\qbar$ is not a complete information equilibrium.

We next present the sufficient and necessary condition under which all fixed points are complete information fixed points. Besides, we derive a sufficient condition on the set of fixed points $\FP$ and the average payoff functions to ensure that the strategy played in the fixed point is equivalent to a complete information equilibrium, although the fixed point belief may not be a complete information belief.

\begin{proposition}\label{prop:complete}
 The fixed point set $\FP = \{\(\thetasran, \qsran\)|\qsran \in \EQ(\thetasran)\}$ if and only if $[\theta] \setminus \Sequiv\(\q\)$ is a non-empty set for any $\theta \in \Delta(\S) \setminus \{\thetasran\}$ and any $\q \in \EQ(\theta)$. 
 
 Furthermore, for a fixed point $\(\thetabar, \qbar\)$, $\qbar = \qsran$ if (i) There exists a positive number $\xi >0$ such that $[\thetabar] \subseteq \Sequiv(\q)$ for any $\|\q- \qbar\|< \xi$; and (ii) The payoff function $\usi(\qi, \qmi)$ is concave in $\qi$ for all $\i \in \I$ and all $\s \in [\thetabar]$. 
\end{proposition}

Proposition \ref{prop:complete} is intuitive: A belief $\theta$ cannot be a fixed point belief if a parameter in its support can be distinguished from the true parameter $\sran$ with an equilibrium corresponding to $\theta$, i.e. $[\theta]\setminus \Sequiv(\q)$ is non-empty for any $\q \in \EQ(\theta)$. Hence, set $\FP$ is comprised of only the complete information fixed points if and only if none of the beliefs in the set $\Delta(\S) \setminus \{\thetasran\}$ is a fixed point belief.


Besides, for any fixed point $\(\thetabar, \qbar\)$, since $\qbar_i$ is a best response strategy of $\qbar_{\mi}$, $\qbar_i$ is a local maximizer of the expected payoff function $\mathbb{E}_{\thetabar}[u_i^s(\qi, \qbar_{\mi})]$. Condition (i) in Proposition \ref{prop:complete} ensures that the value of the expected payoff function is identical to that with the true parameter $\sran$ for any $\qi$ belonging to a small neighborhood of $\qbar_i$. Therefore, $\qbar_i$ must be a local maximizer of the payoff function with the true parameter $u_i^{\sran}(\qi, \qbar_{\mi})$. Moreover, since condition (ii) provides that payoffs are concave functions of $\q_i$, $\qbar_i$ must also be a global maximizer of $u_i^{\sran}(\qi, \qbar_{\mi})$. Thus, any fixed point strategy $\qbar$ is an equilibrium of the game with complete information of $\sran$. 

Next we present three illustrative examples to further discuss the properties of fixed points in our learning dynamics. 

\vspace{0.2cm}
\noindent\textbf{Example 1. (Cournot competition)} A set of $\I$ firms produce an identical product and compete in a market. In each stage $\t$, firm $\i$'s strategy is their production level $\qit\in [0, 3]$. The price of the product is $p^{\t}=\alpha^\s - \beta^s \(\sum_{\i \in \I} \qit\)+ \epsilon^s$, where $\s=\(\alpha^s, \beta^s\)$ is the unknown parameter vector in the price function, and $\epsilon$ is a random variable with zero mean. The set of parameter vectors is $\S=\{s_1, s_2\}$, where $\s_1 = \(2, 1\)$ and $\s_2= \(4, 3\)$. The true parameter is $\sran=s_1$. The marginal cost of each firm is 0. Therefore, the payoff of firm $\i$ in stage $\t$ is $\ct_i=\qit \(\alpha^\s - \beta^\s \(\sum_{\i \in \I} \qit\)+ \epsilon^s\)$ for each $\s \in \S$. The information system updates belief $\thetat$ based on the total production $\sum_{\i \in \I} \qit$ and the realized price $p^\t$.

This game has a potential function, and the best response correspondence $\BR(\q, \theta)$ is a contraction mapping for all $\q \in \Q$ and all $\theta \in \Delta(\S)$. Thus, all three types of best response update rules satisfy Assumption \ref{asu}.\footnote{For any $\theta \in \Delta(\S)$ and any $\q \in \Q$, the best response strategy is $\BR_i(\theta, \q) = \{\frac{\mathbb{E}_{\theta}[\alpha^s]}{2\mathbb{E}_{\theta}[\beta^s]} - \frac{1}{2}\sum_{j \neq i} q_j\}$, where $\mathbb{E}_{\theta}[\alpha^s]=\sum_{\s \in \S}\theta(\s) \alpha^{s}$ and $\mathbb{E}_{\theta}[\beta^s]=\sum_{\s \in \S}\theta(\s) \beta^{s}$. Then, for any $\q, \q' \in \Q$, we have $\|\BR(\theta, \q)- \BR(\theta, \q')\| <\frac{1}{4} \|\q- \q'\|$,
i.e. $\BR(\theta, \q)$ is a contraction mapping. Thus, for each of the three best response dynamics, $\F(\theta, \q)$ is also a contraction mapping. From the Banach's fixed-point theorem, the sequence of strategies converges to the equilibrium strategy in $\EQ(\theta)$ under all three best response dynamics.} From Theorem \ref{theorem:convergence}, the states of the learning dynamics converge to a fixed point with probability 1 with all three types of strategy updates. The complete information fixed point is $\thetasran=\(1,0\)$ and $\qsran=\(2/3, 2/3\)$. Additionally, $\theta^{\dagger}= \(0.5, 0.5\)$ and $\q^{\dagger}=\(0.5, 0.5\) \in \EQ(\theta^{\dagger})$ is also a fixed point since $[\theta^{\dagger}] \subseteq \Sequiv(\q^{\dagger})=\{\s_1, \s_2\}$. Thus, $\(\theta^{\dagger}, \q^{\dagger}\)$ is another fixed point. Note that at $\q^{\dagger}$, the two parameters $s_1$ and $s_2$ lead to identical price distributions, and thus cannot be distinguished.

In fact, since any $\theta \neq \thetasran$ must include $\s_2$ in the support set, one can show that $\q^{\dagger}= \(0.5, 0.5\)$ is the only strategy profile for which $\s_1$ and $\s_2$ are payoff-equivalent. Thus, there does not exist any other fixed points apart from $\(\thetasran, \qsran\)$ and $\(\theta^{\dagger}, \q^{\dagger}\)$; i.e. $\FP=\left\{(\thetasran, \qsran), (\theta^{\dagger}, \q^{\dagger})\right\}$.

\vspace{0.2cm}
\noindent\textbf{Example 2.(Zero sum game)}
  Two players $\i  \in \{1, 2\}$ repeatedly play a zero-sum game with identical convex and closed strategy sets $\Q_1=\Q_2=[0,6]$. For any strategy profile $\q$, the payoff of each player is $\c_1 = - \c_2 = v^\s(\q) + \epsilon^\s$, where 
  \begin{align*}
    v^\s(\q)= \(\max\(|\qt_1 - \qt_2|, \s\) - \s\)^2 - 2(\qt_1)^2,
\end{align*}
and $\s \in \S = \{1, 3, 5\}$ is the unknown parameter. The true parameter $\sran=3$. Belief is updated by an information platform based on the strategy profiles and the realized payoffs. 

All three types of best response update rules satisfy Assumption \ref{asu} in this zero sum game.\footnote{For any $\theta \in \Delta(\S)$, $q_1=0$ maximizes the expected utility of player 1. Thus, regardless of the sequence of player 2's strategies, the sequence of player 1's strategy converges to 0 under all three best response dynamics. Additionally, the sequence of player 2's strategies converges to a best response strategy in $\BR_2(\theta, 0)= \{Q_2|\q_2 \leq \min\{[\theta]\}\}$. Since $\EQ(\theta)=\{\(\q_1, \q_2\)|\q_1=0, \q_2 \leq \min\{[\theta]\}\}$, the sequence of strategies converges to an equilibrium strategy under all three best response dynamics.} From Theorem \ref{theorem:convergence}, the sequence of states converges to a fixed point w.p. 1. 

The set of complete information fixed points is $\thetasran=\(0, 1, 0\)$ and $\EQ(\thetasran)= \{\(\qsran_1, \qsran_2\)|\qsran_1=0, ~ \qsran_2 \leq 3\}$. Apart from the complete information fixed points, any $\theta^{\dagger} \in \Delta(\S) \setminus \{(0,0,1)\}$ and any $\q^{\dagger} \in \{\(\qbar_1, \qbar_2\)|\q^{\dagger}_1=0, ~ \q^{\dagger}_2 \leq \min\{[\theta^{\dagger}]\}\}$ is also a fixed point. This is because for any belief $\theta^{\dagger}$ that assigns positive probability on $\s=1$ or $\s=3$, $\q^{\dagger}_1=0$ and $\q^{\dagger}_2$ such that $|\q^{\dagger}_2-\q^{\dagger}_2|\leq \min\{[\theta^{\dagger}]\}$ is an equilibrium, and the two parameters $\s=1$ and $\s=3$ are payoff equivalent at $\q^{\dagger}$. 

Moreover, we can check that conditions (i) and (ii) in Proposition \ref{prop:complete} are satisfied by any fixed point $\(\theta^{\dagger}, \q^{\dagger}\)$. Thus, any fixed point strategy in the set $\{\(\q^{\dagger}_1, \q^{\dagger}_2\)|\q^{\dagger}_1=0, ~ \q^{\dagger}_2 \leq \min\{[\theta^{\dagger}]\}\}$ is a complete information equilibrium although $\theta^{\dagger}$ is not a complete information belief.

\vspace{0.2cm}
\noindent\textbf{Example 3. (Investment game)}
Two players repeatedly play an investment game. In each stage $\t$, the strategy $\qit \in [0, 1]$ is the non-negative level of investment of player $\i$. Given the strategy profile $\qt = \(\qt_1, \qt_2\)$, the return of a unit investment is randomly realized according to $r^\t=\s +\qt_1+ \qt_2 + \epsilon^\s$, where $\s \in \S=\{0,1,2\}$ is the unknown parameter that represents the average baseline return and $\epsilon^s$ is the noise term. The true parameter is $\sran=1$. The stage cost of investment for each player is $3 \(\qit\)^2$. Therefore, the payoff of each player $\i \in \I$ is $\ct_i= \qit(\s +\qt_1+ \qt_2+\epsilon^\s) - 3 \(\qit\)^2 = \qit(\s -2 \qit+ \qt_{\mi} + \epsilon^\s)$ for all $\s \in \S$. In each stage $\t$, the information system updates belief $\thetat$ based on the total investment $\qt_1+\qt_2$ and the unit investment return $r^\t$. 

This game is a supermodular game, and it is also dominance solvable. All three best response dynamics satisfy Assumption \ref{asu}.\footnote{For any $\theta \in \Delta(\S)$ and any $\q \in \Q$, the best response strategy is $\BR(\theta, \q) = \{\frac{\mathbb{E}_{\theta}[\s] + \q_2}{4}, \frac{\mathbb{E}_{\theta}[\s] + \q_1}{4}\}$, where $\mathbb{E}_{\theta}[\s]=\sum_{\s \in \S}\theta(\s) \s$. Same as Example 1, for any $\q, \q' \in \Q$, we have $\|\BR(\theta, \q)- \BR(\theta, \q')\| = \frac{1}{4} \|\q-\q'\| < \|\q- \q'\|$,
i.e. $\BR(\theta, \q)$ is a contraction mapping. Thus, under any one of the three best response dynamics, $\F(\theta, \q)$ is also a contraction mapping, and the sequence of strategies converges to an equilibrium strategy in $\EQ(\theta)$.} Thus, states converge to a fixed point with probability 1. In this game, since $\Sequiv(\q)=\{\sran=1\}$ for any $\q \in \Q$, the unique fixed point is the complete information fixed point, i.e. $\FP = \{\(\thetasran, \qsran\)= \(\(0, 1, 0\), \(1/3, 1/3\)\)\}$.

%% file: stability.tex
\subsection{Stability}\label{subsec:stability}
In this section, we analyze both global and local stability properties of fixed point belief $\thetabar$ and the associated equilibrium set $\EQ(\thetabar)$. 
\begin{definition}[Global stability]\label{def:global}
A fixed point belief $\thetabar \in \Delta(\S)$ and the associated equilibrium set $\EQ(\thetabar)$ are \emph{globally stable} if for any initial state $\(\thetazero, \q^1\)$, the beliefs of the learning dynamics $\(\thetat\)_{\t=1}^{\infty}$ converge to $\thetabar$ and the strategies $\(\qt\)_{\t=1}^{\infty}$ converge to $\EQ(\thetabar)$ with probability 1. 
\end{definition}
Thus, global stability requires that that the convergent fixed point belief and the corresponding equilibrium set do not depend on the initial state. 

We next introduce the definition of local stability. For any $\epsilon>0$, we define an $\epsilon$-neighborhood of belief $\thetabar$ as $N_{\epsilon}(\thetabar) \deleq \left\{\theta \left\vert \|\theta - \thetabar\|< \epsilon\right. \right\}$. For any $\delta>0$, we define the $\delta$-neighborhood of equilibrium set as $N_{\delta}(\EQ(\thetabar)) \deleq  \left\{ \q \left\vert \dist\(\q, \EQ(\thetabar)\)<\delta\right.\right\}$, where $\dist\(\q, \EQ(\thetabar)\) = \min_{\q'\in \EQ(\thetabar)}\|\q-\q'\|$ is the Euclidean distance between $\q$ and the set $\EQ(\thetabar)$. %

\begin{definition}[Local stability]\label{def:local}
A fixed point belief $\thetabar \in \Delta(\S)$ and the associated equilibrium set $\EQ(\thetabar)$ are \emph{locally stable} if for any $\gamma \in (0,1)$ and any $\thetaepfin, \loadepfin>0$, there exist $\thetaep, \loadep>0$ such that for the learning dynamics that starts with $\thetazero\in \neighinitheta(\thetabar)$ and $\q^1 \in \neighiniload(\EQ(\thetabar))$, the following holds:
\begin{align}\label{eq:local}
\lim_{\t \to \infty} \pro\(\thetat \in \neighinftheta(\thetabar), ~ \qt\in \neighinfload(\EQ(\thetabar))\)>\gamma.\end{align}
\end{definition}

Thus, local stability requires that when the learning starts with an initial state that is sufficiently close to a fixed point belief $\thetabar$ and the associated equilibrium set $\EQ(\thetabar)$, then the sequence of beliefs (resp. sequence of strategies) is guaranteed to be arbitrarily close to $\thetabar$ (resp. $\EQ(\thetabar)$), with arbitrarily high probability. In other words, when the belief $\thetabar$ and the equilibrium strategy set $\EQ(\thetabar)$ are locally stable, the learning dynamics is robust to small perturbations around $\thetabar$ and $\EQ(\bar\theta)$. On the other hand, if $\thetabar$ and $\EQ(\bar\theta)$ are locally unstable, then there exists a non-zero probability $\iota>0$ such that the state of learning dynamics can leave the neighborhood of $\thetabar$ and $\EQ(\thetabar)$ with probability at least $\iota$ even when the initial belief $\thetazero$ (resp. strategy $\q^1$) is arbitrarily close to $\thetabar$ (resp. $\EQ(\thetabar)$).

Note that both global and local stability notions are not defined for a single fixed point, but rather for the
tuple $\(\bar\theta, \EQ(\bar\theta)\)$, i.e. fixed points with an identical belief $\bar\theta$. This is important when the game has multiple equilibria; i.e., $\EQ(\theta)$ is not a singleton set for some belief $\theta\in\Delta(S)$. That is, our stability notions do not hinge on the convergence to a particular equilibrium in the fixed point equilibrium set $\EQ(\thetabar)$.

We provide a necessary and sufficient condition for global stability: 
\begin{proposition}\label{prop:global}
There exists a globally stable fixed point if and only if all fixed points are complete information fixed points, i.e. $\Omega = \left\{\(\thetasran, \EQ(\thetasran)\)\right\}$. In this case, $\(\thetasran, \EQ(\thetasran)\)$ is globally stable.
\end{proposition}

This result is quite intuitive: If the set $\FP$ contains another fixed point that is not a complete information fixed point, then whether the states of learning dynamics converge to the complete information fixed point or another fixed point depends on the initial state; hence no fixed point in the set can be globally stable. Also recall from Proposition \ref{prop:complete} that all fixed points being complete information fixed points is equivalent to the condition that any parameter other than the true parameter $\sran$ can be distinguished from $\sran$ at the equilibrium. From Proposition \ref{prop:global}, we know that this condition is also equivalent to the existence of globally stable fixed points.    






To prove local stability, we assume that the following set of conditions hold: 
\begin{assumption}\label{as:stability}
For a fixed point belief $\thetabar$ and the associated equilibrium set $\EQ(\thetabar)$, $\exists \epsilon, ~\delta >0$ such that the neighborhoods $N_{\epsilon}\(\thetabar\)$ and $N_{\loaddelta}\(\EQ(\thetabar)\)$ satisfy\\
(A2a) \emph{Local upper hemicontinuity}: $\EQ(\theta)$ is upper-hemicontinuous in $\theta$ for any $\theta \in N_{\epsilon}\(\thetabar\)$.\\
(A2b) \emph{Local invariance}: Neighborhood $N_{\loaddelta}(\EQ(\thetabar))$ is a locally invariant set of the best response correspondence, i.e. $\BR(\theta, \q) \subseteq N_{\loaddelta}(\EQ(\thetabar))$ for any $\q \in N_{\loaddelta}\(\EQ(\thetabar)\)$ and any $\theta \in N_{\epsilon}\(\thetabar\)$. \\
(A2c) \emph{Local consistency}: Fixed point belief $\thetabar$ forms a consistent payoff estimate in the local neighborhood $N_\delta(EQ(\bar\theta))$, i.e. $[\bar\theta]\subseteq \Sequiv(q)$ for any $q\in N_\delta(EQ(\bar\theta))$.
\end{assumption}

\begin{theorem}\label{theorem:stability}
A fixed point belief $\bar\theta\in\Delta(S)$ and the associated equilibrium set $EQ(\bar\theta)$ is locally stable under the learning dynamics \eqref{eq:update_belief} and \eqref{eq:generic} if Assumptions \ref{asu} and \ref{as:stability} are satisfied. 
\end{theorem}

From Theorem \ref{theorem:convergence}, we know that Assumption \ref{asu} ensures the convergence of beliefs and strategies under local perturbations. We now discuss the role of each of the three conditions in Assumption 2 towards local stability. Firstly, the local upper hemicontinuity condition \emph{(A2a)} guarantees that the convergent equilibrium strategy remains close to the original fixed point equilibrium when the belief is locally perturbed. Secondly, the local invariance condition \emph{(A2b)} guarantees that the strategy sequence resulting from the strategy updates remains within the local invariant neighborhood of the fixed point equilibrium. We remark that for games with complete information, local invariance reduces to the standard condition on the existence of invariant set for best response strategy updates under no parameter uncertainty, and this property is sufficient to ensure the local stability of complete information equilibrium. Hence, the conditions of local upper hemicontinuity and local invariance conditions together ensure that the strategy sequence in our learning dynamics does not leave the local neighborhood of $\EQ(\thetabar)$ so long as the perturbed beliefs remain close to $\thetabar$.  

Finally, the local consistency condition \emph{(A2c)} ensures that \eqref{eq:update_belief} keeps the beliefs close to $\thetabar$. Under this condition, any parameter in the support of $\thetabar$ remains to be payoff equivalent to $\sran$ for any strategy in a local neighborhood of $\EQ(\thetabar)$. That is, $\thetabar$ forms a consistent estimate of players' payoffs not just at fixed point strategy $\qbar$, but also when the strategy is locally perturbed around $\qbar$. Therefore, the Bayesian belief update keeps the beliefs of all parameters in $[\thetabar]$ close to their respective probabilities in $\thetabar$ when the strategies are in the local neighborhood, and eventually any parameters that are not in $[\thetabar]$ are excluded by the learning dynamics.

We now detail the proof ideas of Theorem \ref{theorem:stability} (the formal proof is given in Appendix \ref{apx:proof}). From Definition \ref{def:local}, to prove local stability, we need to characterize the local neighborhoods $N_{\epsilon^1}\(\thetabar\)$ and $N_{\delta^1}\(\EQ(\thetabar)\)$ of the initial state $\(\thetazero, \q^1\)$ such that \eqref{eq:local} is satisfied. 
In our proof, we first show via Lemma \ref{lemma:constrained_set} that \eqref{eq:local} is satisfied if the sequence of states -- beliefs and strategies -- remain with probability higher than $\gamma$ in the specifically constructed neighborhoods $N_{\epsilonhat}\(\thetabar\)$ and $N_{\delta}\(\EQ(\thetabar)\)$, respectively; here, $\epsilonhat \in (0, \epsilon)$ and $\epsilon$, and $\delta$ are chosen according to Assumption 2.  
Subsequently, in Lemmas \ref{lemma:stopping_time} and \ref{lemma:other_parameters}, we precisely characterize the neighborhoods $N_{\epsilon^1}\(\thetabar\)$ and $N_{\delta^1}\(\EQ(\thetabar)\)$ such that the sequence of beliefs and strategies starting from initial state $(\theta^1,q^1)\in (N_{\epsilon^1} (\bar\theta) \times N_{\delta^1}(\EQ(\bar\theta))$ remains in the respective neighborhoods $N_{\hat \epsilon}(\bar\theta)$ and $N_{\delta}(\EQ(\bar\theta))$ that we specifically construct in Lemma \ref{lemma:constrained_set} with probability higher than $\gamma$.  


%

In Lemma \ref{lemma:constrained_set}, parts \emph{(i)} and \emph{(ii)} show that under Assumption \emph{(A2a)} -- \emph{(A2b)}, the properties of local upper-hemicontinuity and local invariance hold in the neighborhoods $N_{\epsilonhat}\(\thetabar\)$ and $N_{\delta}\(\EQ(\thetabar)\)$. Additionally, part \emph{(iii)} shows that if the belief sequence and strategy sequence are in respective sets $N_{\epsilonhat}(\bar\theta)$ and $N_\delta(EQ(\bar\theta))$, then the convergent state must be in $N_{\thetaepfin}(\thetabar)$ and $N_{\bar{\delta}}\(\EQ(\thetabar)\)$.

\begin{lemma}\label{lemma:constrained_set}
Under Assumptions \ref{asu} and (A2a) -- (A2b),
\begin{enumerate}
    \item[(i)] For any $\loadepfin>0$, $\exists \epsilon' \in \(0, \epsilon\)$ such that any $\theta \in N_{\epsilon'}(\thetabar)$ satisfies $\EQ(\theta) \subseteq N_{\loadepfin}(\EQ(\thetabar))$.
    \item[(ii)] For any $\thetaepfin>0$, $\BR(\theta, \q) \subseteq N_{\delta}(\EQ(\thetabar))$ for all $\q \in N_{\delta}(\EQ(\thetabar))$ and all $\theta \in N_{\epsilonhat}\(\EQ(\thetabar)\)$, where $\epsilonhat = \min\{\epsilon, \epsilon', \thetaepfin\}$.  
    \item[(iii)] $\lim_{\t \to \infty} \pro\(\thetat \in \neighinftheta(\thetabar),\right.$ $ \left. \qt \in  \neighinfload(\EQ(\thetabar))\)\geq \pro\(\thetat \in  N_{\epsilonhat}(\thetabar), ~ \qt \in N_{\delta}(\EQ(\thetabar)), ~\forall \t\)$.
\end{enumerate}
\end{lemma}

In Lemma \ref{lemma:constrained_set}, \emph{(i)} follows from Assumption \emph{(A2a)} that $\EQ(\theta)$ is upper-hemicontinuous in $\theta$ in the local neighborhood $N_{\epsilon}\(\thetabar\)$. Then, we obtain \emph{(ii)} from Assumption \emph{(A2b)} that $N_{\delta}\(\EQ(\thetabar)\)$ is an invariant set of the best response correspondence. Furthermore, if beliefs are in $N_{\epsilonhat}(\thetabar)$ for all stages, then the convergent belief must also be in $N_{\epsilonhat}(\thetabar) \subseteq N_{\thetaepfin}(\thetabar)$. Based on Theorem \ref{theorem:convergence}, the sequence of strategies converges. Since $N_{\epsilonhat}(\thetabar) \subseteq N_{\epsilon'}(\thetabar)$, we know from \emph{(i)} in Lemma \ref{lemma:constrained_set} that the convergent strategy is an equilibrium in the neighborhood $N_{\loadepfin}(\EQ(\thetabar))$. Thus, \emph{(iii)} holds.


Thanks to Lemma \ref{lemma:constrained_set} \emph{(iii)}, to prove local stability as in \eqref{eq:local}, it remains to be established that there exist $N_{\thetaep}(\thetabar)$ and $N_{\loadep}\(\EQ(\thetabar)\)$ for the initial belief $\thetazero$ and strategy $\q^1$ such that $\pro\(\thetat \in  N_{\epsilonhat}(\thetabar),\right.$ $\left. \qt \in N_{\delta}(\EQ(\thetabar)), ~\forall \t\) > \gamma$. In particular, $\thetat \in N_{\epsilonhat}\(\thetabar\)$ is guaranteed if $|\thetat(\s)-\thetabar(\s)|\leq \frac{\epsilonhat}{|\S|}$ for all $\s \in \S$. We separately analyze the beliefs of all $\s \in \S\setminus [\thetabar]$ (i.e. the set of parameters with zero probability in $\thetabar$) in Lemma \ref{lemma:stopping_time}, and that of $\s \in [\thetabar]$ in Lemma \ref{lemma:other_parameters}. Additionally, parts \emph{(a)} and \emph{(b)} in Lemma \ref{lemma:constrained_set} are useful in Lemmas \ref{lemma:stopping_time} and \ref{lemma:other_parameters} for constructing  $\thetaep$ and $\loadep$.


Before proceeding, we need to define the following thresholds:

\begin{subequations}
\begin{align}
    \rhoone &\deleq \min_{\s \in [\thetabar]} \left\{\frac{(1-\gamma) \thetabar(\s)\epsilonhat}{(1-\gamma +|\Sbar|)(|\S\setminus [\thetabar]|+1)|\S|+(1-\gamma) \epsilonhat}\right\}, \label{eq:rhoone} \\
    \rhotwo &\deleq\frac{\epsilonhat}{(|\Sbar|+1)|\S|}, \label{eq:rhotwo}\\
    \rhothree &\deleq \min_{\s \in [\thetabar]}\left\{\frac{\epsilonhat - |\Sbar| |\S|\rhotwo \thetabar(\s)}{|\S|-|\Sbar||\S| \rhotwo},~  \frac{\epsilonhat}{|\S|+ |\Sbar|\(\thetabar(\s)|\S|+ \epsilonhat\)}, ~ \thetabar(\s) \right\}. \label{eq:rho_three}
\end{align}
\end{subequations}

Lemma \ref{lemma:stopping_time} below shows that if the initial belief $\thetazero$ is in the neighborhood $N_{\rhoone}(\thetabar)$, then $\thetat(\s) \leq \rhotwo$ for all $\s \in \S\setminus [\thetabar]$ in all stages of the learning dynamics with probability higher than $\gamma$. Note that $\thetat(\s) \leq \rhotwo$ ensures $|\thetat(\s)-\thetabar(\s)|< \frac{\epsilonhat}{|\S|}$ since $\thetabar(\s)=0$ for all $\s \in \Sbar$ and $\rhotwo < \frac{\epsilonhat}{|\S|}$. Additionally, the threshold $\rhotwo$ is specifically constructed to bound the beliefs of the remaining parameters in $[\thetabar]$, which will be used later in Lemma \ref{lemma:other_parameters}.

\begin{lemma}\label{lemma:stopping_time}
For any $\gamma \in (0, 1)$, if the initial belief satisfies
\begin{subequations}
\begin{align}
    &\thetazero(\s)< \rhoone, \quad \forall \s \in \Sbar, \label{eq:epsilonlow}\\
    &\thetabar(\s) - \rhoone < \thetazero(\s) < \thetabar(\s)+\rhoone, \quad \forall \s \in [\thetabar],\label{eq:epsilonlow_positive}
\end{align}
\end{subequations}
then 
\begin{align}\label{eq:nonsupport}
    \pro\(\thetat(\s)\leq \rhotwo, ~\forall \s \in \Sbar, ~ \forall \t\)>\gamma.
\end{align}
\end{lemma}

In the proof of Lemma \ref{lemma:stopping_time}, we say that the belief $\thetat(\s)$ completes an upcrossing of the interval $[\rhoone, \rhotwo]$ if $\thetat(\s)$ increases from less than $\rhoone$ to higher than $\rhotwo$. Note that if the belief of a parameter $\s \in \Sbar$ is initially smaller than $\rhoone$ but later becomes higher than $\rhotwo$ in some stage $\t$, then the belief sequence $\(\theta^j(\s)\)_{j=1}^{\t}$ must have completed at least one upcrossing of $[\rhoone, \rhotwo]$ before stage $\t$. Therefore, $\thetat(\s)\leq \rhotwo$ for all $\t$ is equivalent to that the number of upcrossings completed by the belief is zero. 

Additionally, by bounding the initial belief of parameters $\s \in [\thetabar]$ as in \eqref{eq:epsilonlow_positive}, we construct another interval $\left[\rhoone/\(\thetabar(\sran)- \rhoone\), \rhotwo\right]$ such that the number of upcrossings with respect to this interval completed by the sequence of belief ratios $\(\frac{\thetat(\s)}{\thetat(\sran)}\)_{\t=1}^{\infty}$ is no less than the number of upcrossings with respect to interval $[\rhoone, \rhotwo]$ completed by $\(\thetat(\s)\)_{\t=1}^{\infty}$. Recall that the sequence of belief ratios $\(\frac{\thetat(\s)}{\thetat(\sran)}\)_{\t=1}^{\infty}$ forms a martingale process (Lemma \ref{lemma:theta}). By applying Doob's upcrossing inequality, we obtain an upper bound on the expected number of upcrossings completed by the belief ratio corresponding to each parameter $\s \in \Sbar$, which is also an upper bound on the expected number of upcrossings made by the belief of $\s$. Using Markov's inequality and the upper bound of the expected number of upcrossings, we show that with probability higher than $\gamma$, no belief $\thetat(\s)$ of any parameter $\s \in \Sbar$ can ever complete a single upcrossing with respect to the interval $[\rhoone, \rhotwo]$ characterized by \eqref{eq:rhoone} -- \eqref{eq:rhotwo}. Hence, $\thetat(\s)$ remains lower than the threshold $\rhotwo$ for all $\s \in \Sbar$ and all $\t$ with probability higher than $\gamma$.

Furthermore, Lemma \ref{lemma:other_parameters} utilizes another set of conditions on the initial belief and strategy; these conditions ensure that the beliefs of the remaining parameters $\s \in [\thetabar]$ satisfy $|\thetat(\s)-\thetabar(\s)|<\frac{\epsilonhat}{|\S|}$, and the strategy $\qt \in N_{\delta}\(\EQ(\thetabar)\)$ for all $\t$ so long as $\thetat(\s)<\rhotwo$ for any parameter $\s \in \S \setminus [\thetabar]$. Recall that $\thetat(\s)<\rhotwo$ for all $\s \in \S\setminus [\thetabar]$ is satisfied with probability higher than $\gamma$ under the conditions provided in Lemma \ref{lemma:stopping_time}.

\begin{lemma}\label{lemma:other_parameters}
Under Assumption (A2b) -- (A2c), if $|\thetazero(\s)-\thetabar(\s)|<\rhothree$ for all $\s \in [\thetabar]$ and $\q^1 \in N_{\delta}\(\EQ(\thetabar)\)$, then 
\begin{align}\label{eq:ensure}
\pro\(\left.\begin{array}{l}
|\thetat(\s)-\thetabar(\s)|<\frac{\epsilonhat}{|\S|}, ~\forall \s \in [\thetabar], ~\forall \t\\
\text{and } \qt \in N_{\delta}\(\EQ(\thetabar)\), ~\forall \t
\end{array}\right\vert
\thetat(\s)<\rhotwo, ~\forall \s \in \Sbar, ~\forall \t\)=1.
\end{align}
\end{lemma}

We prove this lemma by mathematical induction. Since $\rhotwo < \frac{\epsilonhat}{|\S|}$ as in \eqref{eq:rhotwo}, under the condition that $\thetat(\s)<\rhotwo$ for all $\s \in \S\setminus [\thetabar]$ and all $\t$, we know that $\thetat(\s)<\frac{\epsilonhat}{|\S|}$ for all $\s \in \S\setminus [\thetabar]$ and all $\t$. In any stage $\t$, assume that $|\thetat(\s)-\thetabar(\s)|<\frac{\epsilonhat}{|\S|}$ for all $\s \in [\thetabar]$ and $\qt \in N_{\delta}\(\EQ(\thetabar)\)$. Then, $\thetat \in N_{\epsilonhat}\(\thetabar\)$ in stage $\t$. Additionally, under local consistency condition in Assumption \emph{(A2c)}, $\s \in [\thetabar]$ remains to be payoff equivalent at $\qt$. Thus, we can show that the belief of the next stage must satisfy $|\thetat(\s)-\thetabar(\s)|<\frac{\epsilonhat}{|\S|}$ for all $\s \in [\thetabar]$, which ensures that $\thetatone \in N_{\epsilonhat}\(\thetabar\)$. Since $\epsilonhat \leq \epsilon$ as in part \emph{(ii)} of Lemma \ref{lemma:constrained_set}, we know that the updated strategy $\q^{\t+1}$ is in $N_{\delta}\(\EQ(\thetabar)\)$. Hence, we obtain \eqref{eq:ensure} by induction. 

Finally, by setting $\thetaep=\min\{\rho^1, \rho^3\}$ and $\loadep=\delta$, where $\rho^1$, $\rho^3$ are as in \eqref{eq:rhoone}, \eqref{eq:rho_three} and $\delta$ is given by Assumption \ref{as:stability}, the initial state in $N_{\thetaep}(\thetabar)$ and $N_{\delta^1}\(\EQ(\thetabar)\)$ satisfies the conditions in Lemmas \ref{lemma:stopping_time} and \ref{lemma:other_parameters}. Then, by combining \eqref{eq:nonsupport} and \eqref{eq:ensure}, we obtain that all beliefs and strategies are in the neighborhoods $N_{\epsilonhat}\(\thetabar\)$ and $N_{\delta}\(\EQ(\thetabar)\)$ respectively with probability higher than $\gamma$. From (c) in Lemma \ref{lemma:constrained_set}, we know that $\lim_{\t \to \infty} \pro\(\thetat \in \neighinftheta(\thetabar), \qt \in  \neighinfload(\EQ(\thetabar))\) \geq \gamma$. Thus, we have constructed the local neighborhoods of the initial state that satisfy \eqref{eq:local}, and we conclude Theorem \ref{theorem:stability}.

We discuss the local and global stability properties of the fixed points in Examples 1 -- 3. 

\vspace{0.2cm}
\noindent\textbf{Example 1. (continued)} Since the complete information fixed point is not the unique fixed point, no fixed point is globally stable. We now show that the complete information fixed point $\thetasran=(1, 0)$, $\qsran=\(2/3, 2/3\)$ is locally stable. Consider $\epsilon=1/3$ and $\delta=1$. We can check that all three conditions in Assumption \ref{as:stability} are satisfied in the neighborhoods $N_{\epsilon}\(\thetasran\)$ and $N_{\delta}\(\qsran\)$, and thus this fixed point is locally stable. On the other hand, the other fixed point $\theta^{\dagger}=\(0.5, 0.5\)$ and $\q^{\dagger}=\(0.5, 0.5\)$ does not satisfy the local consistency condition since the two parameters $\s_1$ and $\s_2$ can be distinguished when the strategy is perturbed in any local neighborhood of $\q^{\dagger}$.

\vspace{0.2cm}
\noindent\textbf{Example 2. (continued)} Since the complete information fixed point is not the unique fixed point, no fixed point is globally stable. Moreover, by setting $\epsilon=1/2$ and $\delta= 6$, we can check that all fixed points in $\FP$ satisfy the three conditions in Assumption \ref{as:stability}, and thus are locally stable.  

\vspace{0.2cm}
\noindent\textbf{Example 3. (continued)}
The unique fixed point of the public good investment game is the complete information fixed point $\(\thetasran, \qsran\)= \(\(0, 1, 0\), \(1/2, 1/3\)\)$. From Proposition \ref{prop:global}, the complete information fixed point is globally stable.

%% file: MAP.tex
\section{Extensions}\label{sec:variant}
In this section, we consider three types of extensions of the learning model introduced in Sec. \ref{sec:basic_model}: (1) Learning with two timescales; (2) Learning in games with finite strategy set; (3) Learning with maximum a posteriori probability (MAP) or ordinary least squares (OLS) estimates. 

\vspace{0.2cm}

\noindent\textbf{(1) Learning with two timescales.} Consider the case where strategy update is at a faster timescale compared with the belief updates, i.e. $\lim_{t \to \infty}\t_{t+1}-\kt = \infty$ with probability 1. Under Assumption \ref{asu}, as $t \rightarrow \infty$, the strategies between two belief updates $\kt$ and $\t_{t+1}$ converge to an equilibrium strategy profile in $\EQ\(\theta^{\kt}\)$ before the next belief update in stage $\t_{t+1}$. Then, the updated belief $\theta^{\t_{t+1}}$ forms an accurate payoff estimate given the equilibrium strategy. Our convergence result (Theorem \ref{theorem:convergence}) holds for this two timescale dynamics. The local and global stability results in Theorem \ref{theorem:stability} and Proposition \ref{prop:global} also hold in an analogous manner. 

\vspace{0.2cm}

\input{extension}

\vspace{0.2cm}
\noindent\textbf{(3) MAP and OLS estimates.} Now consider a continuous and bounded parameter set $\S$, and that the initial belief $\thetazero(\s)$ is a probability density function of $\s$ on the set $\S$, and $\thetazero(\s)>0$ for all $\s \in \S$. Since the unknown parameter $\s$ is continuous, Bayesian belief update in \eqref{eq:update_belief} at stage $\t_{t+1}$ is as follows: 
\begin{align*}
\theta^{\t_{t+1}}(\s)&= \frac{\theta^{\kt}(\s)\prod_{\t=\kt}^{\t_{t+1}-1}\phibar^\s(\ct|\qt)}{\int_{s \in \S} \theta^{\kt}(\s)\prod_{\t=\kt}^{\t_{t+1}-1} \phibar^{\s}(\ct|\qt) ds}, \quad \forall \s \in \S.
\end{align*}
 Instead of computing the full posterior belief in each stage (which entails computing the continuous integration in the denominator of the Bayesian update), we consider learning with maximum a posteriori (MAP) estimator that maximizes the posterior belief of the unknown parameter:
\begin{align}\tag{$\theta_M$-update}\label{eq:map}
\theta^{\t_{t+1}}_M(\s)&=\argmax_{\s \in \S}\theta^{\t_{t+1}}(\s)= \argmax_{\s \in \S}\theta^{\kt}(\s)\prod_{\t=\kt}^{\t_{t+1}-1}\phibar^\s(\ct|\qt). 
\end{align}
Note that if the initial belief $\thetazero$ is a uniform distribution of all parameters, then the MAP estimate is also a maximum likelihood estimate (MLE). 
 
Our result on convergence of state (Theorem 1) can be directly extended to this case of learning with MAP estimate. In particular, under Assumption \ref{asu}, the sequence of MAP estimates converges to a payoff equivalent parameter $\thetabar_M \in \Sequiv(\qbar)$ given the fixed point strategy profile, and the strategies converge to an equilibrium strategy $\qbar \in \EQ(\thetabar_M)$ of game $\G$ with parameter $\thetabar_M$. 

Moreover, under Assumptions \ref{asu} and \ref{as:stability}, we can check that if the initial belief $\thetazero$ is in a small local neighborhood of the belief vector that assigns probability 1 to a fixed point MAP estimate $\thetabar_M$ and the strategy profile is in a small local neighborhood of the equilibrium $\EQ(\thetabar_M)$, then the convergent belief remains in a small neighborhood of the singleton belief vector so that the MAP estimate remains to be $\thetabar_M$ and the equilibrium set is $\EQ(\thetabar_M)$. Therefore, analogous to Theorem \ref{theorem:stability}, we can conclude that $\(\thetabar_M, \EQ(\thetabar_M)\)$ is locally stable under conditions given by Assumptions \ref{asu} and \ref{as:stability}.   

 Finally, we consider a special case, where the average payoff functions are affine in strategies: 
\begin{align}\label{eq:linear_utility}
    \c_i=\(\q, 1\) \cdot \s_i+\epsilon_i^\s, \quad \forall \i \in I.
\end{align} 
The unknown parameter vector is $\s=\(\s_i\)_{\i \in \I}$, where $\s_i$ has $|\q|+1$ dimensions. The noise term $\epsilon_i^\s$ is realized from a normal distribution with zero mean and finite variance.

From stage $1$ to $\kt$, player $\i$'s realized payoff $\(\c_i^{
\t}\)_{\t=1}^{\kt}$ can be written as a linear function of the strategies $\(\qt\)_{\t=1}^{\kt}$ in the following matrix form: 
\begin{align*}
    \underbrace{\left(\begin{array}{l}
    \c_i^1\\
    \c_i^2\\
    \vdots\\
    \c_i^{\kt}
    \end{array}\right)}_{Y_i^{\kt}} = \underbrace{\left(\begin{array}{ll}
    \q^1, &1\\
    \q^2, &1\\
    \vdots &\vdots\\
    \q^{\kt},& 1
    \end{array}\right)}_{\widetilde{\Q}^{\kt}} \s_i+\left(\begin{array}{l}\epsilon_i^1\\
    \epsilon_i^2\\
    \vdots\\
    \epsilon_i^{\kt}
    \end{array}
    \right).
\end{align*}
The OLS estimate is $\shat^{\kt}=\(\shat_i^{\kt}\)_{\i \in \I}$ where
\begin{align}\label{eq:OLS}\tag{$\shat$ - update}
    \shat_i^{\kt}= \(\(\widetilde{\Q}^{\kt}\)' \widetilde{\Q}^{\kt}\)^{-1}\(\widetilde{\Q}^{\kt}\)'Y_i^{\kt}, \quad \forall \i \in \I, \quad \forall \kt.
\end{align}

In learning dynamics with OLS estimates, the convergence of the OLS estimates can be viewed as a special case of learning with MAP estimator because $\shat^{\kt}$ is identical to the MLE estimator $\theta^{\kt}_M$ when each player's payoff as in \eqref{eq:linear_utility} is an affine function of the strategy profile plus a noise term with Normal distribution. Therefore, we obtain the same convergence result in the learning with OLS estimate as in learning with MAP estimate. That is, the OLS estimates converge to an estimate $\sbar \in \S$ such that  $u^{\sbar}_i(\qbar) =u^{\sran}_i(\qbar)$ for all $\i \in \I$, and strategies converge to $\qbar \in \EQ(\sbar)$ with probability 1. Furthermore, as we have shown in Example 1, when payoff functions are linear in players' strategies, only the complete information fixed point satisfies the locally consistency condition Assumption \emph{(A2c)}. Thus, no other fixed point satisfies the sufficient conditions for local stability. 

%% file: extension.tex
\noindent\textbf{(2) Learning in Games with Finite Strategy Set.}
Our results in Sec. \ref{sec:main} can be extended to learning in games where strategy sets are finite and players can choose mixed strategies. In this game, each player $\i$'s action set (pure strategies) is a finite set $\Ai$, and the action profile (pure strategy profile) is denoted as $\a=\(\ai\)_{\i \in \I} \in A= \prod_{\i \in \I} \Ai$. Given any parameter $\s$ and any action profile $\a$, the distribution of players' payoff $\c$ is $\phi^\s\(\c|\a\)$. 

We denote player $\i$'s mixed strategy as $\qi = \(\qi(\ai)\)_{\ai \in \Ai} \in \Qi = \Delta\(\Ai\)$, where $\qi(\ai)$ is the probability of choosing the action $\ai$. The strategy set $\Qi$ is bounded and convex. Players' action profile in each stage $\t$, denoted as $\actiont =\(\ai^{\t}\)_{\i \in \I}$, is realized from the mixed strategy profile $\qt$.

Analogous to \eqref{eq:update_belief}, the information system updates the belief $\theta^{\kt}$ based on actions $\(\a^{\t}\)_{\t=\kt}^{\t_{t+1}-1}$ and the realized payoff vectors $\(\ct\)_{\t=\kt}^{\t_{t+1}-1}$ as follows: 
\begin{align*}
\theta^{k_{t+1}}(\s)&= \frac{\theta^{\kt}(\s)\prod_{\t=\kt}^{\t_{t+1}-1}\phibar^\s(\ct|\actiont)}{\sum_{s' \in \S} \theta^{\kt}(\s') \prod_{\t=\kt}^{\t_{t+1}-1}\phibar^{\s'}(\ct|\actiont)}, \quad \forall \s \in \S.
\end{align*}
Similar to Sec. \ref{sec:basic_model}, we consider three types of best response updates: 
\begin{enumerate}
    \item \emph{Simultaneous best response dynamics.} All players choose an action that is best response to the updated belief and their opponents' action profile: 
    \begin{align*}
       \a^{\t+1}_{\i}& \in \BR_i(\thetatone, \a_{\mi}^{\t}), \quad \forall \i \in \I.  
    \end{align*}
    \item \emph{Sequential best response dynamics.} Players change their actions to be a best response strategy of other opponents' actions one by one:  
        \begin{align*}
        \a_i^{\t+1}\left\{\begin{array}{ll} \in \BR_i(\thetatone, \at_{\mi}), &\quad \text{if } \t ~ mod ~|\I| = \i.  \\
        =\ait,& \quad \text{otherwise}.  
        \end{array}
        \right.
    \end{align*}
    \item \emph{Fictitious play.} The mixed strategy $\qit$ represents player $\i$'s empirical frequency of actions in previous stages $1, \dots, \t$. In each stage $\t$, all players best respond to their opponents' empirical frequency $\qtmi$: 
    \begin{align*}
    \ait \in \BR_i(\thetat, \qtmi), \quad 
    \qi^{\t+1}=\frac{\t}{\t+1}\qit+ \frac{1}{\t+1} \ait, \quad \forall \i \in \I, \quad \forall \t.
\end{align*}
\end{enumerate}

We extend the definition of payoff equivalent parameters in Definition \ref{def:payoff_equivalence} as follows: Parameter $\s$ is payoff-equivalent to $\sran$ given $\q \in \Q$ if the distribution of payoffs under $\s$ is identical to that under $\sran$ for all actions that are assigned with positive probability given $\q$. Therefore, the payoff-equivalent parameter set given $\q$ is defined as $\Sequiv(\q) \deleq \left\{\S\left\vert D_{KL}\(\phi^{\s*}\(\c|\n\)||\phi^{\s}\(\c|\n\)\)=0, ~ \forall \n \in [\q]\right.\right\}$, where $[\q]=\{\N|\q(\n)>0\}$ is the support set of the mixed strategy profile $\q$. 

The convergence result in Theorem \ref{theorem:convergence} can be readily extended to games with finite strategy sets: Under Assumption \ref{asu}, the beliefs $\(\thetat\)_{\t =1}^{\infty}$ converge to a fixed point belief $\thetabar$ that accurately estimates the payoff distribution for all action profiles that are taken with positive probability, and the strategies $\(\qt\)_{\t=1}^{\infty}$ converge to the equilibrium set $\EQ(\thetabar)$. 

The results on global and local stability properties in Proposition \ref{prop:global} and Theorem \ref{theorem:stability} also hold for games with finite strategy set. Moreover, for games with a finite strategy set, any fixed point that satisfies the sufficient condition of local stability must be a complete information fixed point. This is because any local perturbation of a fixed point strategy profile can lead to a mixed strategy with full support on all action profiles, and these mixed strategies can distinguish any parameter $\s \neq \sran$ from $\sran$. Thus, local consistency condition in Assumption \emph{(A2c)} is only satisfied by the complete information belief $\thetasran$. 

\vspace{0.2cm}
\noindent\textbf{Example 4.} Atomic routing games with unknown cost parameters (\cite{wu2020bayesian}) \\
A set of players $\I$ repeatedly travel in a transportation network with edge set $\E$. Each player is associated with an origin-destination pair, and their action set $\Ai$ is the set of paths -- sequences of edges -- that connect their origin to their destination. In each stage $\t$, given an action profile $\at = \(\ait\)_{\i \in \I}$, the load on each edge $x^{\t}_e$ is the total number of players using that edge. The network has an uncertain state $\s \in \S$ that affects the costs of edges. Given any $\s$, the cost of edge $\e$ is $c_e^{\t}= d_e^s(x_e^{\t})+ \epsilon_e^s$, where $d_e^s(\cdot)$ is an increasing function of $x_e^{\t}$ and $\epsilon_e^s$ is a noise term with zero mean. The cost of each player is $\ct_i =  \sum_{\e \in \ait} c_e^{\t}$. The traffic information system updates the belief of the unknown parameter vector $\s$ based on the load vector $x^{\t}=\(x_{e}^{\t}\)_{\e \in \E}$ of all edges, and the realized travel time cost of all edges that are taken $c^{\t}=\(c^{\t}_e\)_{\e \in \{\E|x_e^{\t}>0\}}$. 

The states in our learning dynamics converge to a fixed point under both sequential best response dynamics and fictitious play. On the other hand, the simultaneous best response dynamics can be cyclic.\footnote{For any $\theta \in \Delta(\S)$, the routing game has a potential function. Therefore, the sequence of strategies under \eqref{eq:br} and \eqref{eq:br_diminish} converges to an equilibrium strategy in $\EQ(\theta)$ (\cite{monderer1996potential, monderer1996fictitious, marden2009joint}). However, the sequence of strategies under \eqref{eq:sbr} may not converge. For example, consider a two route network $\{e_1, e_2\}$ with identical cost functions $\mathbb{E}_{\theta}[d_e^s(x_e)] = \mathbb{E}_{\theta}[\s]x_e+1$, where $\s>0$ is the unknown parameter. There are two players. If learning that starts with both players taking $e_1$, then under \eqref{eq:sbr}, both players take $e_2$ in all even stages 2, 4, $\dots$, and take $e_1$ in all odd stages 1, 2, $\dots$. Therefore, the strategy sequence does not converge under \eqref{eq:sbr}, and Assumption \ref{asu} is not satisfied.} At the fixed point, the belief $\thetabar$ accurately estimates the costs of all edges that are taken by travelers in $\qbar$, and all players take the route that minimizes their expected costs given $\thetabar$. However, the fixed point strategy $\qbar$ may not be the complete information equilibrium because $\thetabar$ can form wrong estimate on edges that are not taken in $\qbar$. Finally, the complete information fixed point is locally stable. 

%% file: conclusion.tex
\section{Concluding Remarks}
In this article, we studied stochastic learning dynamics induced by a set of strategic players who repeatedly play a game with an unknown parameter. We analyzed the convergence of beliefs and strategies induced by the stochastic dynamics, and derived conditions for local and global stability of fixed points. We also provided a simple condition which guarantees the convergence of strategies to complete information equilibrium. 

A future research question of interest is to analyze the learning dynamics when players seek to efficiently learn the true parameter by choosing off-equilibrium strategies. When there are one or more parameters that are payoff equivalent to the true parameter at fixed point, complete learning requires players to take strategies that may reduce their individual payoffs in some stages. In our setup, if a player were to choose a non-equilibrium strategy, the information resulting from that player's realized payoff would be incorporated into the belief update, and the new belief is known to all players. Under what scenarios the utility-maximizing players will choose their strategies to engage such explorative behavior is an interesting question, and worthy of further investigation. 

Another promising extension is to study multi-agent reinforcement learning problem from a Bayesian viewpoint. In such settings, the unknown parameter changes over time according to a Markovian transition process, and players may have imperfect or no knowledge of the underlying transition kernel. The ideas presented in this article are useful to analyze how players learn the belief estimates of payoffs that depend on the latent Markov state, and adaptively adjust their strategies that converges to an equilibrium. 

%% file: proof_convergence.tex
\noindent\textbf{\emph{Proof of Lemma \ref{lemma:theta}.}} \\
First, we show that for any parameter $\s \in \S$, the sequence $\(\frac{\theta^{\kt}(\s)}{\theta^{\kt}(\sran)}\)_{t=1}^{\infty}$ is a non-negative martingale, and hence converges with probability 1. Note that for any $t=1, 2, \dots $, and any parameter $\s \in \S \setminus \{\sran\}$, we have the following from \eqref{eq:update_belief}: 
\begin{align*}
\frac{\theta^{\t_{t+1}}(\s)}{\theta^{\t_{t+1}}(\sran)}= \frac{\theta^{\kt}(\s) \cdot \prod_{\t=\kt}^{\t_{t+1}-1}\phibar^{\s}(\ct|\qt)}{\theta^{\kt}(\sran) \cdot \prod_{\t=\kt}^{\t_{t+1}-1}\phibar^{\sran}(\ct|\qt)}.
\end{align*}
Now starting from any initial belief $\thetazero$, consider a sequence of strategies $Q^{\kt-1}\deleq \(\qj\)_{j=1}^{\kt-1}$ and a sequence of realized payoffs $Y^{\kt-1}\deleq \(\cj\)_{\j =1}^{\kt-1}$ before stage $\kt$. Then, the expected value of $\frac{\theta^{\t_{t+1}}(\s)}{\theta^{\t_{t+1}}(\sran)}$ conditioned on $\thetazero$, $Q^{\kt-1}$ and $Y^{\kt-1}$ is as follows: 
\begin{align}\label{conditional_iterate}
\mathbb{E}\left[\left.\frac{\theta^{\t_{t+1}}(\s)}{\theta^{\t_{t+1}}(\sran)}\right\vert \thetazero, Q^{\kt-1}, Y^{\kt-1}\right]&= \frac{\theta^{\kt}(\s)}{\theta^{\kt}(\sran)} \cdot \mathbb{E}\left[ \frac{\prod_{\t=\kt}^{\t_{t+1}-1}\phibar^{\s}(\ct|\qt)}{ \prod_{\t=\kt}^{\t_{t+1}-1}\phibar^{\sran}(\ct|\qt)} \right]
\end{align}
where $\theta^{\kt}$ is the repeatedly updated belief from $\thetazero$ based on $Q^{\kt-1}$ and $Y^{\kt-1}$ using \eqref{eq:update_belief}. Note that
\begin{align*}
&\mathbb{E}\left[ \frac{\prod_{\t=\kt}^{\t_{t+1}-1}\phibar^{\s}(\ct|\qt)}{ \prod_{\t=\kt}^{\t_{t+1}-1}\phibar^{\sran}(\ct|\qt)}\right]= \int_{y^{\kt}y^{\kt+1}y^{\t_{t+1}-1}}  \(\frac{\prod_{\t=\kt}^{\t_{t+1}-1}\phibar^{\s}(\ct|\qt)}{ \prod_{\t=\kt}^{\t_{t+1}-1}\phibar^{\sran}(\ct|\qt)}\) \cdot  \(\prod_{\t=\kt}^{\t_{t+1}-1}\phibar^{\sran}(\ct|\qt) \)d y^{\kt}y^{\kt+1}y^{\t_{t+1}-1}\\
=&\int_{y^{\kt}y^{\kt+1}y^{\t_{t+1}-1}}  \prod_{\t=\kt}^{\t_{t+1}-1}\phibar^{\s}(\ct|\qt) d y^{\kt}y^{\kt+1}y^{\t_{t+1}-1} =1.
\end{align*}
Hence, for any $\t=1, 2, \dots$,  
\begin{align*}
\mathbb{E}\left[\left.\frac{\theta^{\t_{t+1}}(\s)}{\theta^{\t_{t+1}}(\sran)}\right\vert \thetazero, Q^{\kt-1}, Y^{\kt-1}\right]&= \frac{\theta^{\kt}(\s)}{\theta^{\kt}(\sran)}, \quad \forall \s \in \S.
\end{align*}
Again, from \eqref{eq:update_belief} we know that $\frac{\theta^{\kt}(\s)}{\theta^{\kt}(\sran)}\geq 0$. Hence, the sequence $\(\frac{\theta^{\kt}(\s)}{\theta^{\kt}(\sran)}\)_{t=1}^{\infty}$ is a non-negative martingale for any $\s \in \S$. From the martingale convergence theorem, we conclude that $\frac{\theta^{\kt}(\s)}{\theta^{\kt}(\sran)}$ converges with probability 1. 

Next we show that the sequence $\left(\log\theta^{\kt}(\sran)\right)_{t=1}^{\infty}$ is a submartingale, and hence converges with probability 1. We define the estimated density function of payoffs $\(\ct\)_{\t=\kt}^{k_{t+1}-1}$ with belief $\theta$ as $\mu \(\(\ct\)_{\t=\kt}^{\t_{t+1}-1}\left\vert\theta^{\kt}, \(\qt\)_{\t=\kt}^{\t_{t+1}-1}\right.\)\deleq\sum_{\s \in \S} \theta^{\kt}(\s)\prod_{\t=\kt}^{\t_{t+1}-1}\phibar^{s}(\ct|\qt)$. From \eqref{eq:update_belief}, we have:
\begin{align*}
    &\mathbb{E}\left[\left.\log\theta^{\t_{t+1}}(\sran)\right\vert \thetazero, Q^{\kt-1}, Y^{\kt-1}\right]= \mathbb{E}\left[\left.\log\(\frac{\theta^{\kt}(\sran)\prod_{\t=\kt}^{\t_{t+1}-1}\phibar^{\sran}(\ct|\qt)}{\mu \(\(\ct\)_{\t=\kt}^{\t_{t+1}-1}\left\vert\theta^{\kt}, \(\qt\)_{\t=\kt}^{\t_{t+1}-1}\right.\)}\)\right\vert \thetazero, Q^{\kt-1}, Y^{\kt-1}\right]\\
  &= \log\theta^{\kt}(\sran) +\mathbb{E}\left[\log\(\frac{\prod_{\t=\kt}^{\t_{t+1}-1}\phibar^{\sran}(\ct|\qt)}{\mu \(\(\ct\)_{\t=\kt}^{\t_{t+1}-1}\left\vert\theta^{\kt}, \(\qt\)_{\t=\kt}^{\t_{t+1}-1}\right.\)}\)\right]\\
  & =\log\theta^{\kt}(\sran) +\int_{y^{\kt}y^{\kt+1}y^{\t_{t+1}-1}} \(\prod_{\t=\kt}^{\t_{t+1}-1}\phibar^{\sran}(\ct|\qt)\) \log\(\frac{\prod_{\t=\kt}^{\t_{t+1}-1}\phibar^{\sran}(\ct|\qt)}{\mu \(\(\ct\)_{\t=\kt}^{\t_{t+1}-1}\left\vert\theta^{\kt}, \(\qt\)_{\t=\kt}^{\t_{t+1}-1}\right.\)}\) d y^{\kt}y^{\kt+1}y^{\t_{t+1}-1} \\
    &= \log\theta^{\kt}(\sran)+  D_{KL}\(\prod_{\t=\kt}^{\t_{t+1}-1}\phibar^{\sran}(\ct|\qt)\left\vert \left\vert\mu \(\(\ct\)_{\t=\kt}^{\t_{t+1}-1}\left\vert\theta^{\kt}, \(\qt\)_{\t=\kt}^{\t_{t+1}-1}\right.\)\right.\right. \) \geq \log\theta^{\kt}(\sran),
\end{align*}
where the last inequality is due to the non-negativity of KL divergence between $\prod_{\t=\kt}^{\t_{t+1}-1}\phibar^{\sran}(\ct|\qt)$ and $\mu \(\(\ct\)_{\t=\kt}^{\t_{t+1}-1}\left\vert\theta^{\kt}, \(\qt\)_{\t=\kt}^{\t_{t+1}-1}\right.\)$. Therefore, the sequence $\left(\log\theta^{\kt}(\sran)\right)_{t=1}^{\infty}$ is a submartingale. Additionally, since $\log\theta^{\kt}(\sran)$ is bounded above by zero, by the martingale convergence theorem $\log\theta^{\kt}(\sran)$ converges with probability 1. Hence, $\theta^{\kt}(\sran)$ must also converge with probability 1.

From the convergence of $\frac{\theta^{\kt}(\s)}{\theta^{\kt}(\sran)}$ and $\theta^{\kt}(\sran)$, we conclude that $\theta^{\kt}(\s)$ converges with probability 1 for any $\s \in \S$. Since for any $\t=\kt+1, \dots, \t_{t+1}-1$, $\thetat= \theta^{\kt}$, we know that $\thetat$ also converges. Let the convergent vector be denoted as $\thetabar= \(\thetabar(\s)\)_{\s \in \S}$. We can check that for any $\t$, $\thetat(\s) \geq 0$ for all $\s \in \S$ and $\sum_{\s \in \S}\thetat(\s)=1$. Hence, $\thetabar$ must satisfy $\thetabar(\s) \geq0$ for all $\s \in \S$ and $\sum_{\s \in \S}\thetabar(\s)=1$, i.e. $\thetabar$ is a feasible belief vector. \QEDA

Before proceeding, we show that the best response correspondence is upper hemicontinuous in the belief and the strategy profile. This result follows directly from the Berge's theorem of maximum and the fact that the expected utility function $\mathbb{E}_{\theta}\left[u_i^s(\qi, \qmi)\right]$ is continuous in $\theta$ and $\q$. 
\begin{lemma}\label{lemma:uh}
For any $\theta \in \Delta(\S)$, any $\i \in \I$ and any $\qmi \in \Qmi$, $\BR(\theta, \qmi)$ is upper-hemicontinuous in $\theta$ and $\qmi$. 
\end{lemma}


We are now ready to prove Lemma \ref{lemma:q}. 
\vspace{0.2cm}

\noindent\emph{Proof of Lemma \ref{lemma:q}.} 
For any stage $\K \geq 1$, we construct an auxiliary sequence of strategies $\(\qhatt\)_{\t=1}^{\infty}$ as follows: First, we set $\qhat^{\t}=\q^{\t}$ for all $\t=1, \dots, \K$. Then, for any $\t> K$, we define the following subsequences: 
\begin{itemize}
\item[-] We define $\tilde{\q}^{\t+1} \deleq \argmin_{\tilde{\q}\in F(\thetabar, \q^{\t})} \|\tilde{\q} - \q^{\t+1}\|$. That is, $\tilde{\q}^{\t+1}$ is a strategy updated from $\qt$ with the fixed point belief $\thetabar$ (i.e. $\tilde{\q}^{\t+1} \in F(\thetabar, \q^{\t})$). Additionally, $\tilde{\q}^{\t+1}$ is the closest to $\q^{\t+1}$ -- the strategy in stage $\t+1$ of the original sequence -- among all strategies in the set $F(\thetabar, \q^{\t})$. 
\item[-] We define the auxiliary strategy $\qhat^{\t+1}\deleq \argmin_{\q\in F(\thetabar, \qhat^{\t})} \|\q- \tilde{q}^{\t+1}\|$. That is, $\qhat^{\t+1}$ is a strategy updated from $\qhat^{\t}$ with the fixed point belief $\thetabar$ (i.e. $\qhat^{\t+1} \in F(\thetabar, \qhat^{\t})$). Additionally, $\qhat^{\t+1}$ is the closest to $\tilde{q}^{\t+1}$ among all strategies in the set $F(\thetabar, \qhat^{\t})$.
\end{itemize}Therefore, for any $\t > \K$, we have: 
\begin{align}\label{eq:dist}
    \|\q^{\t+1}-\tilde{\q}^{\t+1}\|= \dist\(\q^{\t+1}, F(\thetabar, \qt)\),~ \|\tilde{\q}^{\t+1} -\qhat^{\t+1}\|= \dist\(\tilde{q}^{\t+1}, F(\thetabar, \qhatt)\).
\end{align}
We next show by mathematical induction that for any $\ell\geq 1$, $\lim_{\K \to \infty} \|\q^{\K+\ell}- \qhat^{\K+\ell}\| =0$. To begin with, for $\ell=1$, we have
\begin{align}
    &\|\q^{\K+1}- \qhat^{\K+1}\| \leq \|\q^{\K+1}- \tilde{q}^{\K+1}\|+ \|\tilde{\q}^{\K+1} - \qhat^{\K+1}\| \notag \\
    \stackrel{\eqref{eq:dist}}{=}&\dist\(\q^{K+1}, F\(\thetabar, \q^{\K}\)\) + \dist\(\tilde{q}^{\K+1}, F\(\thetabar, \qhat^{\K}\)\). \label{eq:Kl}
\end{align}
Since $\thetat$ converges to $\thetabar$ (Lemma \ref{lemma:theta}), $F(\theta, \q)$ is upper hemicontinuous in $\theta$ (Lemma \ref{lemma:uh}), and $\q^{K+1} \in F(\theta^{K+1}, \q^K)$, we know that $\lim_{\K \to \infty}\dist\(\q^{K+1}, F\(\thetabar, \q^{\K}\)\)=0$. Additionally, since $\qhat^K=\q^K$ and $\tilde{\q}^{K+1} \in F(\thetabar, \q^{\K})$, $\dist\(\tilde{q}^{\K+1}, F\(\thetabar, \qhat^{\K}\)\)=0$. Therefore, $\lim_{\K \to \infty} \|\q^{\K+1}- \qhat^{\K+1}\| =0$.

Now, assume that $\lim_{\K \to \infty} \|\q^{\K+\ell}- \qhat^{\K+\ell}\| =0$ for some $\ell\geq 1$, we need to prove that $\lim_{\K \to \infty} \|\q^{\K+\ell+1}- \qhat^{\K+\ell+1}\| =0$. Similar to \eqref{eq:Kl}, we have
\begin{align*}
  \|\q^{\K+\ell+1}- \qhat^{\K+\ell+1}\| \leq \dist\(\q^{K+\ell+1}, F\(\thetabar, \q^{\K+\ell}\)\) + \dist\(\tilde{q}^{\K+\ell+1}, F\(\thetabar, \qhat^{\K+\ell}\)\)
\end{align*}
Analogous to $\ell=1$, since $F(\theta, \q)$ is upper hemicontinuous in $\theta$, $\lim_{\K \to \infty} \dist\(\q^{K+\ell+1}, F\(\thetabar, \q^{\K+\ell}\)\) =0$. Additionally, since $\lim_{\K \to \infty} \|\q^{\K+\ell}- \qhat^{\K+\ell}\| =0$, $F(\theta, \q)$ is upper hemicontinuous in $\q$, and $\tilde{q}^{\K+\ell+1} \in F\(\thetabar, \q^{\K+\ell}\)$, we know that $\lim_{\K \to \infty}  \dist\(\tilde{q}^{\K+\ell+1}, F\(\thetabar, \qhat^{\K+\ell}\)\)=0$.  Therefore, we have $\lim_{\K \to \infty} \|\q^{\K+\ell+1}- \qhat^{\K+\ell+1}\| =0$. By mathematical induction, we conclude that for any $\ell\geq 1$, $\lim_{\K \to \infty} \|\q^{\K+\ell}- \qhat^{\K+\ell}\|=0$. 

Finally, Assumption \ref{asu} ensures that the strategy update with constant beliefs $\thetabar$ converges to an equilibrium strategy $\qbar \in \EQ(\thetabar)$. That is, for any $\K \geq 1$, $\lim_{\ell \to \infty} \|\qhat^{\K+\ell} - \qbar\|=0$. Therefore, 
\begin{align*}
    \lim_{\t \to \infty} \|\qt-\qbar\| = \lim_{\ell \to \infty} \lim_{\K \to \infty} \|\q^{\K+\ell}-\qbar\| \leq  \lim_{\ell \to \infty} \lim_{\K \to \infty} \|\q^{\K+\ell}-\qhat^{\K+\ell}\| +\lim_{\ell \to \infty} \lim_{\K \to \infty} \|\qhat^{\K+\ell}- \qbar\|=0. 
\end{align*}
Thus, $\lim_{\t \to \infty} \qt=\qbar$. \QEDA

\vspace{0.2cm}

\noindent\emph{\textbf{Proof of Lemma \ref{lemma:consistency}.}} By iteratively applying the belief update in \eqref{eq:update_belief}, we can write:
\begin{align}\label{eq:sequence_update}
    \theta^{\kt}(\s)= \frac{\theta^1(\s) \prod_{\j=1}^{\kt-1}\phibar^{s}(\cj|\qj)}{\sum_{\s' \in \S} \theta^1(\s') \prod_{\j=1}^{\kt-1}\phibar^{s'}(\cj|\qj)}, \quad \forall \s \in \S.
\end{align}
We define $\Phi^\s(\Costhis|\Loadhis)$ as the probability density function of the history of the realized payoffs $\Costhis=\(\c^j\)_{j=1}^{\kt-1}$ conditioned on the history of strategies $\Loadhis= \(\q^j\)_{j=1}^{\kt-1}$ prior to stage $\kt$, i.e. $\Phi^\s(\Costhis|\Loadhis) \deleq \prod_{\j=1}^{\kt-1}\phibar^{s}(\cj|\q^j)$. We rewrite \eqref{eq:sequence_update} as follows: 
\begin{align}
    \theta^{\kt}(\s)&= \frac{\theta^1(\s) \Phi^\s(\Costhis|\Loadhis)}{\sum_{\s' \in \S} \theta^1(\s') \Phi^{\s'}(\Costhis|\Loadhis)} \leq \frac{\theta^1(\s) \Phi^\s(\Costhis|\Loadhis)}{ \theta^1(\s) \Phi^{\s}(\Costhis|\Loadhis)+ \theta^1(\sran) \Phi^{\sran}(\Costhis|\Loadhis)}\notag\\
    &=\frac{\theta^1(\s) \frac{\Phi^\s(\Costhis|\Loadhis)}{\Phi^{\sran}(\Costhis|\Loadhis)}}{ \theta^1(\s) \frac{\Phi^{\s}(\Costhis|\Loadhis)}{\Phi^{\sran}(\Costhis|\Loadhis)}+ \theta^1(\sran)}.\label{eq:bound_theta}
\end{align}
For any $\s \in \S \setminus \Sequiv(\qbar)$, if we can show that the ratio $ \frac{\Phi^\s(\Costhis|\Loadhis)}{\Phi^{\sran}(\Costhis|\Loadhis)}$ converges to 0, then $\theta^{\kt}(\s)$ must also converge to 0. Now, we need to consider two cases:\\
\noindent\emph{Case 1:} $\phi^{\sran}(\cbar|\qbar) \ll \phi^{\s}(\cbar|\qbar)$: In this case, the log-likelihood ratio can be written as: 
\begin{align}\label{eq:log-likelihood}
    \log \(\frac{\Phi^\s(\Costhis|\Loadhis)}{\Phi^{\sran}(\Costhis|\Loadhis)}\)=   \sum_{\j=1}^{\kt-1} \log\(\frac{\phibar^{\s}(\cj|\qj)}{\phibar^{\sran}(\cj|\qj)}\).
\end{align}
For any $\s \in \S$, since $\phibar^{s}(\cj|\qj)$ is continuous in $\q^{j}$, the probability density function of $\etaj$ is also continuous in $\q^{j}$. In Lemma \ref{lemma:q}, we proved that $\(\qt\)_{\t=1}^{\infty}$ converges to $\qbar$. Then, the distribution of $\etaj$ must converge to the distribution of $\log\(\frac{\phibar^\s(\cbar|\qbar)}{\phibar^{\sran}(\cbar|\qbar)}\)$. Note that for any $ \s \in \S \setminus \Sequiv(\qbar)$, the expectation of $\etabar$ can be written as: 
\begin{align*}
\mathbb{E}\left[\etabar\right]= \int_{\cbar}\phibar^{\sran}(\cbar|\qbar) \cdot  \log\(\frac{\phibar^\s(\cbar|\qbar)}{\phibar^{\sran}(\cbar|\qbar)}\)  d \cbar=-D_{KL}\(\phibar^{\sran}(\cbar|\qbar)||\phibar^\s(\cbar|\qbar)\)<0.
\end{align*}  
If we can show that the equation \eqref{eq:hold} below holds, then we can conclude that the log-likelihood sequence defined by~\eqref{eq:log-likelihood} converges to $-\infty$; this would in turn imply that the sequence of likelihood ratios $\frac{\Phi^\s(\Costhis|\Loadhis)}{\Phi^{\sran}(\Costhis|\Loadhis)}$ must converge to 0. But first we need to show: 
\begin{align}\label{eq:hold}
\lim_{t \to \infty}\frac{1}{\kt-1} \log \(\frac{\Phi^\s(\Costhis|\Loadhis)}{\Phi^{\sran}(\Costhis|\Loadhis)}\)&=\lim_{t \to \infty}\frac{1}{\kt-1} \sum_{j=1}^{\kt-1} \etaj \notag\\
&= \mathbb{E}\left[\etabar\right], \quad w.p.~1.
\end{align}

We denote the cumulative distribution function of $\etaj$ as $F^{j}(z): \mathbb{R} \to [0,1]$, i.e. $F^{j}(z)=\pro\(\etaj \leq z\)$. The cumulative distribution function of $\etabar$ is denoted $\bar{F}(z): \mathbb{R} \to [0,1]$, i.e. $\bar{F}(z)=\pro\(\etabar \leq z\)$. Then, 
\begin{align}\label{ind}
    \lim_{j \to \infty}F^j(z)=\bar{F}(z), \quad \forall z \in \mathbb{R}.
\end{align}
For any sequence of realized payoffs $(\cj)_{j=1}^\infty$, we define a sequence of random variables $\Delta =\(\Delta^{j}\)_{j=1}^{\infty}$, where $\Delta^{j}= F^j\(\etaj\)$. Then, we must have $\Delta^{j} \in [0,1]$, and for any $\delta \in [0,1]$, $\pro(\Delta^{j} \leq \delta)=\pro\(F^j\(\etaj\)\leq\delta\)=\delta$. That is, $\Delta^j$ is independently and uniformly distributed on $[0,1]$. Consider another sequence of random variables $\(\Chibarj\)_{j=1}^{\infty}$, where $\Chibarj\deleq \(\bar{F}\)^{-1}(\Delta^j)$. Since $\Delta^j$ is i.i.d. with uniform distribution, $\Chibarj$ is also i.i.d. with the same distribution as $\etabar$. Additionally, since each $\Delta^j$ is generated from the realized payoff $\cj$, $\(\Chibarj\)_{j=1}^{\infty}$ is in the same probability space as $\etaj$. From \eqref{ind}, we know that as $j \to \infty$, $F^j$ converges to $\bar{F}$. Therefore, with probability 1,
\begin{align*}
    \lim_{j\to \infty} \left\vert\etaj-\Chibarj\right\vert=\lim_{j \to \infty}\left\vert\etaj-(\bar{F})^{-1}F^{j}\(\etaj\)\right\vert=0.
\end{align*}
Consequently, with probability 1, 
\begin{align}\label{eq:close}
\lim_{t \to \infty}\left \vert\frac{1}{\kt-1} \sum_{j=1}^{\kt-1}\( \etaj-  \Chibarj\)\right\vert \leq \lim_{t \to \infty}\frac{1}{\kt-1}  \sum_{j=1}^{\kt-1}\left\vert\etaj-\Chibarj\right\vert=0. 
\end{align}
Since $\(\Chibarj\)_{j=1}^{\infty}$ is independently and identically distributed according to the distribution of $\etabar$, from strong law of large numbers, we have:
\begin{align*}
 \lim_{t \to \infty} \frac{1}{\kt-1}\sum_{j=1}^{\kt-1}\Chibarj = \mathbb{E}\left[\etabar\right]=-D_{KL}\(\phibar^{\sran}(\cbar|\qbar)||\phibar^\s(\cbar|\qbar)\), \quad w.p.~1.
\end{align*}

From \eqref{eq:close}, we obtain the following: 
\begin{align}\label{eq:sum_chi}
\lim_{t \to \infty} \frac{1}{\kt-1} \sum_{j=1}^{\kt-1} \etaj= \lim_{t \to \infty} \frac{1}{\kt-1}\sum_{j=1}^{\kt-1}\Chibarj=-D_{KL}\(\phibar^{\sran}(\cbar|\qbar)||\phibar^\s(\cbar|\qbar)\), \quad w.p.~1
\end{align}
Hence, \eqref{eq:hold} holds. Then, for any $\s \in \S \setminus \Sequiv(\qbar)$, $\lim_{t \to \infty} \frac{\Phi^\s(\Costhis|\Loadhis)}{\Phi^{\sran}(\Costhis|\Loadhis)}=0$. Thus, from \eqref{eq:bound_theta}, we know that $\lim_{t \to \infty} \theta^{\kt}(\s)=0$ for all $\s \in \S \setminus \Sequiv(\qbar)$. Since for any $\t=\kt+1, \dots, \t_{t+1}-1$, $\thetat= \theta^{\kt}$, we know that $\lim_{\t \to \infty} \frac{\Phi^\s(Y^{\t-1}|Q^{\t-1})}{\Phi^{\sran}(Y^{\t-1}|Q^{\t-1})}=0$ and $\lim_{\t \to \infty} \thetat(\s)=0$ for all $\s \in \S \setminus \Sequiv(\qbar)$.

Finally, since $\thetazero(\s)>0$ for all $\s \in \S$, the true parameter $\sran$ is never excluded from the belief. Therefore, $\lim_{\t \to \infty} \frac{1}{\t} \log\(\thetat(\sran)\)=0$. For any $\s \in \S \setminus \Sequiv(\qbar)$, we have the following:
\begin{align*}
    &\lim_{\t \to \infty} \frac{1}{\t}\log\(\thetat(\s)\)=   \lim_{\t \to \infty} \frac{1}{\t} \log\(\thetat(\sran)\) +\lim_{\t \to \infty} \frac{1}{\t}\log\(\frac{\thetat(\s)}{\thetat(\sran)}\)\\
    &=\lim_{\t \to \infty} \frac{1}{\t}\log\(\frac{\thetat(\s)}{\thetat(\sran)}\) = \lim_{\t \to \infty} \frac{1}{\t}\log \(\frac{\theta^1(\s)}{\theta^1(\sran)}\)+\lim_{\t \to \infty} \frac{1}{\t} \log\(\frac{\Phi^\s(Y^{\t-1}|Q^{\t-1})}{\Phi^{\sran}(Y^{\t-1}|Q^{\t-1})}\)\\
    &= \mathbb{E} \left[\log\(\frac{\phibar^{\s}(\cbar|\qbar)}{\phibar^{\sran}(\cbar|\qbar)}\)\right]
    =-D_{KL}\(\phibar^{\sran}(\cbar|\qbar)||\phibar^\s(\cbar|\qbar)\), \quad w.p.~1. 
\end{align*}

\noindent\emph{Case 2:} $\phi^{\sran}(\cbar|\qbar)$ is not absolutely continuous in $\phi^{\s}(\cbar|\qbar)$. \\
In this case, $\phi^\s(\cbar|\qbar)=0$ does not imply $\phi^{\sran}(\cbar|\qbar)=0$ with probability 1, i.e. $\pro\(\phi^\s(\cbar|\qbar)=0\)>0$, where $\pro\(\cdot\)$ is the probability of $\cbar$ with respect to the true distribution $\phi^{\sran}(\cbar|\qbar)$. Since the distributions $\phi^\s(\cbar|\q)$ and $\phi^{\sran}(\cbar|\q)$ are continuous in $\q$, the probability $\pro\(\phi^\s(\cbar|\q)=0\)$ must also be continuous in $\q$. Therefore, for any $\epsilon \in \(0, \pro\(\phi^\s(\cbar|\qbar)=0\)\)$, there exists $\delta>0$ such that $\pro\(\phi^\s(\c|\q)=0\) > \epsilon$ for all $\q \in \{\q|\|\q-\qbar\|<\delta\}$. 

From Lemma \ref{lemma:q}, we know that $\lim_{\t \to \infty} \qt = \qbar$. Hence, we can find a positive number $K_1>0$ such that for any $\t > K_1$, $\|\qt - \qbar\|<\delta$, and hence $\pro\(\phi^\s(\ct|\qt)=0\) > \epsilon$. We then have $\sum_{\t=1}^{\infty}\pro\(\phi^\s(\ct|\qt)=0\) = \infty$. Moreover, since the event $\phi^\s(\ct|\qt)=0$ is independent from the event $\phi^\s(\c^{\t'}|\q^{\t'})=0$ for any $\t, \t'$, we can conclude that $\pro\(\phi^\s(\ct|\qt)=0, \text{infinitely often}\)=1$ based on the second Borel-Cantelli lemma. Hence, $\pro\(\phi^\s(\ct|\qt)>0, ~\forall \t\)=0$. From the Bayesian update \eqref{eq:update_belief}, we know that if $\phi^\s(\ct|\qt)=0$ for some stage $\t$, then any belief of $\s$ updated after stage $\t$ is 0. Therefore, we can conclude that $\pro\(\thetat(\s)>0, ~\forall \t\)=0$ with probability 1, i.e. there exists a positive number $\Tstar > K_1$ with probability 1 such that $\thetat(\s)=0$ for any $\t > \Tstar$. 
    \QEDA
    
\vspace{0.2cm}
\noindent\emph{\textbf{Proof of Proposition \ref{prop:complete}.}} 
Firstly, if $[\theta] \setminus \Sequiv(\q)$ is a non-empty set for any $\theta \in \Delta(\S)\setminus \{\thetasran\}$ and any $\q \in \EQ(\theta)$, then no belief with imperfect information $\theta \in \Delta(\S)\setminus \{\thetasran\}$ satisfies \eqref{eq:exclude_distinguished}. That is, only the complete information belief vector $\thetasran$ can be a fixed point belief. Therefore, all fixed point must be complete information fixed points. 

On the other hand, assume for the sake of contradiction that there exists a belief $\theta^{\dagger} \in \Delta(\S) \setminus \{\thetasran\}$ such that $[\theta^{\dagger}] \subseteq \Sequiv(\q^{\dagger})$ for an equilibrium strategy $\q^{\dagger} \in \EQ(\theta^{\dagger})$, then $\(\theta^{\dagger}, \q^{\dagger}\)$, which is not a complete information fixed point, is in the set $\FP$. Thus, we arrive at a contradiction.

Secondly, from condition \emph{(i)} that $[\thetabar] \subseteq \Sequiv(\q)$ for any $\|\q- \qbar\| < \xi$, we have: 
\begin{align}\label{eq:expected}
    \mathbb{E}_{\thetabar}[u_i^{\s}(\q)] = u_i^{\sran}(\q), \quad \forall \i \in \I.
\end{align}
For any $\qbar \in \EQ(\thetabar)$, from condition \emph{(ii)} that $\qbar_i$ is a best response to $\qbar_{\mi}$, $\qbar_i$ must be a local maximizer of $\mathbb{E}_{\thetabar}[u_i^{\s}(\qi, \qbar_{\mi})]$. From \eqref{eq:expected}, $\qbar_i$ is a local maximizer of $u_i^{\sran}(\qi, \qbar_{\mi})$. Since the function $u_i^{\sran}(\qi, \qbar_{\mi})$ is concave in $\qi$, $\qbar_i$ is also a global maximizer of $u_i^{\sran}(\qi, \qbar_{\mi})$, and hence is a best response of $\qbar_{\mi}$ with complete information of $\sran$. Since this argument holds for all $\i \in \I$, $\qbar$ is a complete information equilibrium.\QEDA

%% file: proof_stability.tex
\vspace{0.2cm}
\noindent\textbf{\emph{Proof of Proposition \ref{prop:global}.}}
On one hand, if $\FP = \{\(\thetasran, \EQ(\thetasran)\)\}$, then for any initial state, the learning dynamics converges to a complete information fixed point with belief $\thetasran$ and strategy in $\EQ(\thetasran)$. That is, $\(\thetasran, \EQ(\thetasran)\)$ is globally stable. On the other hand, if there exists another fixed point $\(\theta^{\dagger}, \q^{\dagger}\) \in \FP \setminus \{\(\thetasran, \EQ(\thetasran)\)\}$, then learning that starts with the initial belief $\thetazero =\theta^{\dagger}$ (resp. $\thetazero =\thetasran$) and strategy $\q^1 = \q^{\dagger}$ (resp. $\q^1 = \qsran$) remains at $\(\theta^{\dagger}, \q^{\dagger}\)$ (resp. $\(\thetasran, \qsran\)$) for all stages w.p. 1. Thus, in this case, globally stable fixed points do not exist. \QEDA
\vspace{0.2cm}

\noindent\textbf{\emph{Proof of Lemma \ref{lemma:constrained_set}.}} 
\begin{itemize}
    \item[(i)] From Assumption \emph{(A2a)}, we know that such $\epsilon'$ must exist. 
    \item[(ii)] Since $\epsilonhat \leq \epsilon$, we know from Assumption \emph{(A2b)} that $\BR(\theta, \q) \subseteq N_{\delta}\(\EQ(\thetabar)\)$ for any $\theta \in N_{\epsilonhat}\(\thetabar\)$ and any $\q \in N_{\delta}\(\EQ(\thetabar)\)$. 
    \item[(iii)] Under Assumption \ref{asu}, we know from Theorem \ref{theorem:convergence} that the sequence of the beliefs and strategies converges to a fixed point $\(\theta^{\dagger}, \q^{\dagger}\)$.  If $\thetat \in N_{\epsilonhat}(\thetabar)$ for all $\t$, then $\lim_{\t \to \infty} \thetat = \theta^{\dagger} \in N_{\epsilonhat}(\thetabar) \subseteq N_{\thetaepfin}(\thetabar)$. Additionally, from \emph{(i)} and the fact that $\epsilonhat \leq \epsilon'$, we know that $\lim_{\t \to \infty} \qt = \q^{\dagger} \in \EQ(\theta^{\dagger})\subseteq  N_{\loadepfin}\(\EQ(\thetabar)\)$. Therefore, 
    \begin{align*}
        \lim_{\t \to \infty} \pro\(\thetat \in \neighinftheta(\thetabar) , ~ \qt \in  \neighinfload(\EQ(\thetabar))\)\geq &\pro\(\thetat \in  N_{\epsilonhat}(\thetabar), ~\forall \t\) \\
        \geq &\pro\(\thetat \in  N_{\epsilonhat}(\thetabar), \qt \in N_{\delta}(\EQ(\thetabar)), ~\forall \t\).
    \end{align*}
\end{itemize}
 \QEDA

In the proofs of Lemmas  \ref{lemma:stopping_time} -- \ref{lemma:other_parameters}, we denote $\(\thetatilt\)_{\t=1}^{\infty}$ as an auxiliary belief sequence that is updated in every stage (instead of just updated at $\(\kt\)_{k=1}^{\infty}$). That is, 
\begin{align}\label{eq:update_belief_always}
    \tilde{\theta}^1= \thetazero,\text{ and } ~\tilde{\theta}^{\t+1}(\s) = \frac{\thetatilt(\s)\phi^s(\ct|\qt)}{\sum_{\s' \in \S}\thetatilt(\s')\phi^{s'}(\ct|\qt)}, \quad \forall \s \in \S, \quad \forall \t=1, 2, \dots
\end{align}
From \eqref{eq:update_belief}, we know that 
\begin{align}\label{eq:relationship}
    \thetat=\left\{\begin{array}{ll}
    \thetatilt, &\quad  \text{if  }\t=\kt, \quad \forall k=1, 2, \dots, \\
    \theta^{\t-1}, & \quad \text{otherwise}.
    \end{array}
    \right.
\end{align}
\noindent \textbf{\emph{Proof of Lemma \ref{lemma:stopping_time}.}} First, note that $0< \rhoone < \rhotwo< \frac{\epsilonhat}{|\S|}$. For any $\s \in \Sbar$ and any $\t>1$, we denote $U^{\t}(\s)$ the number of upcrossings of the interval $[\rhoone, \rhotwo]$ that the belief $\thetatil^j(\s)$ completes by stage $\t$. That is, $U^{\t}(\s)$ is the maximum number of intervals $\([\underline{\t}_{i}, \overline{\t}_{i}]\)_{\i=1}^{U^{\t}(\s)}$ with $1 \leq \underline{\t}_{1} < \overline{\t}_{1} < \underline{\t}_{2} <  \overline{\t}_{2}< \cdots< \underline{\t}_{U^{\t}(\s)}< \overline{\t}_{U^{\t}(\s)} \leq \t$, such that $\thetatil^{\underline{\t}_{i}}(\s)<\rhoone< \rhotwo <\thetatil^{\overline{\t}_{i}}(\s)$ for $i=1, \dots U^{\t}(\s)$. Since the beliefs $\(\thetatil^\j(\s)\)_{\j=1}^{\t}$ are updated based on randomly realized payoffs $\(\c^j\)_{j=1}^{\t}$ as in \eqref{eq:update_belief_always}, $U^{\t}(\s)$ is also a random variable. For any $\t>1$, $U^{\t}(\s)\geq1$ if and only if $\thetatil^1(\s)<\rhoone$ and there exists a stage $j \leq \t$ such that $\thetatil^j(\s)>\rhotwo$. Equivalently, $\lim_{\t \to \infty} U^\t(\s)\geq1$ if and only if $\thetatil^1(\s)<\rhoone$ and there exists a stage $\t>1$ such that $\thetatilt(\s)>\rhotwo$. Therefore, if $\thetatil^1(\s)< \rhoone$ for all $\s \in \Sbar$, then:
\begin{align}
    &\pro\(\thetatilt(\s)\leq\rhotwo, ~ \forall \s \in \Sbar, ~\forall \t\) = 1- \pro\(\exists \s \in \Sbar \text{ and } \t, ~s.t. ~\thetatilt(\s)>\rhotwo\) \notag\\
    \geq & 1- \sum_{\s \in \Sbar} \pro\(\exists \t, ~s.t. ~~\thetatilt(\s)>\rhotwo\) = 1- \sum_{\s \in 
    \Sbar}\lim_{\t \to \infty} \pro\(U^\t(\s)\geq 1\). \label{eq:up_one}
\end{align}

Next, we define $\alpha \deleq \thetabar(\sran)-\rhoone$. Since $0<\rhoone < \min_{\s \in [\thetabar]} \{\thetabar(\s)\}$ and $\sran$ is in the support set, we have $\alpha \in (0, \thetabar(\sran))$. If $\thetatil^1(\s)$ satisfies \eqref{eq:epsilonlow} -- \eqref{eq:epsilonlow_positive}, then $\frac{\thetatil^1(\s)}{\thetatil^1(\sran)}< \frac{\rhoone}{\alpha}$ for all $\s \in \Sbar$. Additionally, for any stage $\t$ and any $\s \in \Sbar$, if $\thetatilt(\s)>\rhotwo$, then $\frac{\thetatilt(\s)}{\thetatilt(\sran)} \geq \rhotwo$ because $\thetatilt(\sran)\leq 1$. Hence, whenever $\thetatilt(\s)$ completes an upcrossing of the interval $\left[\rhoone, \rhotwo\right]$, $\frac{\thetatilt(\s)}{\thetatilt(\sran)}$ must also have completed an upcrosssing of the interval $\left[\frac{\rhoone}{\alpha}, \rhotwo\right]$. From \eqref{eq:rhoone} -- \eqref{eq:rhotwo}, we can check that $\frac{\rhoone}{\alpha} < \rhotwo$ so that the interval $\left[\frac{\rhoone}{\alpha}, \rhotwo\right]$ is valid. We denote $\Uhat^\t(\s)$ as the number of upcrossings of the sequence $\(\frac{\thetatil^j(\s)}{\thetatil^j(\sran)}\)_{j=1}^{\t}$ with respect to the interval $\left[\frac{\rhoone}{\alpha}, \rhotwo\right]$ until stage $\t$. Then, $U^\t(\s) \leq \Uhat^\t(\s)$ for all $\t$. Therefore, we can write:
\begin{align}\label{eq:up_two}
    \pro\(U^\t(\s)\geq 1\) \leq \pro\(\Uhat^\t(\s)\geq 1\) \leq \mathbb{E}\left[\Uhat^\t(\s)\right], 
\end{align}
where the last inequality is due to Makov inequality. 

From the proof of Lemma \ref{lemma:theta}, we know that the sequence $\(\frac{\thetatilt(\s)}{\thetatilt(\sran)}\)_{\t=1}^{\infty}$ is a martingale. Therefore, we can apply the Doob's upcrossing inequality as follows: 
\begin{align}\label{eq:up_three}
    \mathbb{E}\left[\Uhat^\t(\s)\right] \leq \frac{\mathbb{E}\left[\max \{\frac{\rhoone}{\alpha}-\frac{\thetatilt(\s)}{\thetatilt(\sran)}, 0\}\right]}{\rhotwo- \frac{\rhoone}{\alpha}} \leq \frac{\frac{\rhoone}{\alpha}}{\rhotwo- \frac{\rhoone}{\alpha}}, \quad \forall \t.
\end{align}
From \eqref{eq:relationship}, \eqref{eq:up_one} -- \eqref{eq:up_three}, and \eqref{eq:rhoone} -- \eqref{eq:rhotwo}, we can conclude that: 
\begin{align*}
    &\pro\(\thetat(\s)\leq\rhotwo, ~ \forall \s \in \Sbar, ~ \forall \t\) =  \pro\(\thetatilt(\s)\leq\rhotwo, ~ \forall \s \in \Sbar, ~ \forall \t\) \\
    \geq &1- \frac{\frac{\rhoone}{\alpha} |\Sbar|}{\rhotwo- \frac{\rhoone}{\alpha}}= 1- \frac{\frac{\rhoone}{\thetabar(\sran)-\rhoone} |\Sbar|}{\rhotwo- \frac{\rhoone}{\thetabar(\sran)-\rhoone}}> \gamma.
\end{align*} \QEDA

\vspace{0.2cm}
\noindent \emph{\textbf{Proof of Lemma \ref{lemma:other_parameters}.}}
From Assumption \emph{(A2c)}, we know that $[\thetabar] \subseteq \Sequiv(\q^1)$ if $\q^1 \in N_{\delta}\(\EQ(\thetabar)\)$. Hence, $\phibar^\s(\c^1|\q^1) = \phibar^{\sran}(\c^1|\q^1)$ for any $\s \in [\thetabar]$ and any realized payoff $\c^1$. Therefore, 
\begin{align}\label{eq:two_one}
    \frac{\thetatil^{2}(\s)}{\thetatil^{2}(\sran)}=\frac{\thetatil^{1}(\s)}{\thetatil^{1}(\sran)}\frac{\phibar^{\s}(\c^1|\q^1)}{\phibar^{\sran}(\c^1|\q^1)}= \frac{\thetatil^{1}(\s)}{\thetatil^{1}(\sran)}, \quad w.p.~1, \quad \forall \s \in [\thetabar].
\end{align}
This implies that $\frac{\sum_{\s \in [\thetabar]} \thetatil^{2}(\s)}{\thetatil^{2}(\sran)}=\frac{\sum_{\s \in [\thetabar]} \thetatil^1(\s)}{\thetatil^1(\sran)}$, and for all $\s \in [\thetabar]$:
\begin{align*}
\frac{\thetatil^{2}(\s)}{\sum_{\s \in [\thetabar]} \thetatil^{2}(\s)}=\frac{\thetatil^{2}(\s)}{\thetatil^{2}(\sran)} \frac{\thetatil^{2}(\sran)}{\sum_{\s \in [\thetabar]} \thetatil^{2}(\s)}=\frac{\thetatil^1(\s)}{\thetatil^1(\sran)} \frac{\thetatil^1(\sran)}{\sum_{\s \in [\thetabar]} \thetatil^1(\s)}=\frac{\thetatil^1(\s)}{\sum_{\s \in[\thetabar]} \thetatil^1(\s)}.
\end{align*}
Thus, we have
\begin{align*}
    \frac{\thetatil^{2}(\s)}{\thetatil^1(\s)}=\frac{\sum_{\s \in [\thetabar]} \thetatil^{2}(\s)}{\sum_{\s \in[\thetabar]} \thetatil^1(\s)}, \quad w.p.~1,  \quad \forall \s \in [\thetabar].\end{align*}

Since $\sum_{\s \in [\thetabar]} \thetatil^1(\s) \leq 1$, if $\thetatil^{2}(\s)<\rhotwo$ for all $\s \in \Sbar$, then we have $\frac{\thetatil^{2}(\s)}{\thetatil^1(\s)}> 1-|\Sbar|\rhotwo$. Additionally, since $\sum_{\s \in [\thetabar]} \thetatil^{2}(\s)<1$ and $\thetatil^1(\s)< \rhothree$ for all $\s \in [\thetabar]$, we have $\frac{\thetatil^{2}(\s)}{\thetatil^1(\s)}<\frac{1}{1-|\Sbar| \rhothree}$. Since by \eqref{eq:rho_three}, $\rhothree \leq  \thetabar(\s)$ for all $\s \in \Sbar$, any $\thetatil^1(\s) \in \(\thetabar(\s)-\rhothree, \thetabar(\s)+ \rhothree\)$ is a non-negative number for all $\s \in [\thetabar]$. Therefore, we have the following bounds: 
\begin{align}\label{eq:refer_end}
\(\thetabar(\s)-\rhothree\) \(1-|\Sbar|\rhotwo\)<\thetatil^{2}(\s) <\frac{\thetabar(\s)+ \rhothree}{1-|\Sbar| \rhothree}.
\end{align}
Since
\begin{align}\label{eq:up}
    \rhothree \stackrel{\eqref{eq:rho_three}}{\leq} \frac{\epsilonhat - |\Sbar| |\S|\rhotwo \thetabar(\s)}{|\S|-|\Sbar||\S| \rhotwo}, \quad \forall \s \in [\thetabar],
\end{align}
we can check that $\(\thetabar(\s)-\rhothree\) \(1-|\Sbar|\rhotwo\) \geq \thetabar(\s) -\frac{\epsilonhat}{|\S|}$ for all $\s \in [\thetabar]$. To ensure the right-hand-side of \eqref{eq:up} is positive, we need to have $\rhotwo < \frac{\epsilonhat}{|\Sbar| |\S| \thetabar(\s)}$ for all $\s \in [\thetabar]$, which is satisfied by \eqref{eq:rhotwo}.  Also, since $\rhothree \stackrel{\eqref{eq:rho_three}}{\leq} \frac{\epsilonhat}{|\S|+ |\Sbar|\(\thetabar(\s)|\S|+ \epsilonhat\)}$ for all $\s \in [\thetabar]$, we have $\frac{\thetabar(\s)+ \rhothree}{1-|\Sbar| \rhothree}< \thetabar(\s)+\frac{\epsilonhat}{|\S|}$ for all $\s \in [\thetabar]$. Therefore, we can conclude that $\thetatil^{2}(\s) \in \(\thetabar(\s)-\frac{\epsilonhat}{|\S|}, \thetabar(\s)+\frac{\epsilonhat}{|\S|}\)$ for all $\s \in [\thetabar]$. Additionally, if $\thetatil^2(\s) \leq \rhotwo < \frac{\epsilonhat}{|\S|}$ for all $\s \in \S \setminus [\thetabar]$, then $\thetatil^2 \in N_{\epsilonhat}\(\thetabar\)$. From \eqref{eq:relationship}, we have $\theta^2 \in N_{\epsilonhat}\(\thetabar\)$. From \emph{(ii)} in Lemma \ref{lemma:constrained_set}, we know that $\BR(\theta^2,\q^1) \in N_{\delta}\(\EQ(\thetabar)\)$. Since $\q^1 \in N_{\delta}\(\EQ(\thetabar)\)$, the updated strategy $\q^2$ given by \eqref{eq:generic} must also be in the neighborhood $N_{\delta}\(\EQ(\thetabar)\)$.

We now use mathematical induction to prove that the belief of any $\s \in [\thetabar]$ satisfies $\thetatil^{\t}(\s) \in \(\thetabar(\s)-\frac{\epsilonhat}{|\S|}, \thetabar(\s)+\frac{\epsilonhat}{|\S|}\)$ for stages $\t>2$. If in stages $j=1, \dots, \t$, $|\thetatil^j(\s)-\thetabar(\s)|<\frac{\epsilonhat}{|\S|}$ for all $\s \in [\thetabar]$ and $\thetatil^j(\s)<\rhotwo<\frac{\epsilonhat}{|\S|}$ for all $\s \in \Sbar$, then $\thetatil^j \in N_{\epsilonhat}\(\thetabar\)$ for all $j=1, \dots, \t$. Thus, from \eqref{eq:relationship} and \emph{(ii)} in Lemma \ref{lemma:constrained_set}, we have $\theta^j \in N_{\epsilonhat}\(\thetabar\)$ and $\BR\(\theta^j, \q^{j-1}\) \subseteq N_{\delta}\(\EQ(\thetabar)\)$. Therefore, $\qj \in N_{\delta}\(\EQ(\thetabar)\)$ for all $j=2, \dots, \t$. 

From Assumption \emph{(A2c)}, we know that  $[\thetabar] \subseteq \Sequiv(\qj)$ for all $j=1, \dots, \t$. Therefore, for any $\s \in [\thetabar]$ and any $j=1, \dots, \t$, $\phibar^\s(\cj|\qj)=\phibar^{\sran}(\cj|\qj)$ with probability 1. Then, by iteratively applying \eqref{eq:two_one}, we have $\frac{\thetatil^{\t+1}(\s)}{\thetatil^1(\s)}=\frac{\sum_{\s \in [\thetabar]} \thetatil^{\t+1}(\s)}{\sum_{\s \in[\thetabar]} \thetatil^1(\s)}$ for all $\s \in [\thetabar]$ with probability 1. Analogous to $\t=2$, we can prove that $|\thetatil^{\t+1}(\s)-\thetabar(\s)|<\frac{\epsilonhat}{|\S|}$ for all $\s \in [\thetabar]$. From \eqref{eq:relationship}, we also have $|\theta^{\t+1}(\s)-\thetabar(\s)|<\frac{\epsilonhat}{|\S|}$ for all $\s \in [\thetabar]$. From the principle of mathematical induction, we conclude that in all stages $\t$,  $|\thetat(\s)-\thetabar(\s)|< \frac{\epsilonhat}{|\S|}$ for all $\s \in [\thetabar]$, and $\qt \in N_{\delta}\(\EQ(\thetabar)\)$ for all $\t$. Therefore, we have proved \eqref{eq:ensure}.

\QEDA

\vspace{0.2cm}

Finally, we are ready to prove Theorem \ref{theorem:stability}. 

\vspace{0.2cm}
\noindent\textbf{\emph{Proof of Theorem \ref{theorem:stability}.}}
We combine Lemmas \ref{lemma:constrained_set} -- \ref{lemma:other_parameters}. For any $\gamma \in (0, 1)$, and any $\thetaepfin, \loadepfin >0$, consider $\loadep =\delta$ and $\thetaep \deleq \min\{\rhoone, \rhothree\}$ given by \eqref{eq:rhoone} and \eqref{eq:rho_three}. If $\thetazero \in N_{\epsilon^1}(\thetabar)$, then $|\thetazero(\s)-\thetabar(\s)|<\thetaep$ for all $\s \in \S$. Recall from \emph{(iii)} in Lemma \ref{lemma:constrained_set}, $\lim_{\t \to \infty} \pro\(\thetat\in N_{\thetaepfin}(\theta), ~ \qt \in \neighinfload(\EQ(\thetabar)) \) \geq  \pro\(\thetat \in N_{\epsilonhat}(\thetabar), ~ \qt \in N_{\delta}\(\EQ(\thetabar)\), ~ \forall \t\)$. Since $\rhotwo \leq \epsilonhat/|\S|$, we further have: 
\begin{align*}
    &\pro\(\thetat \in N_{\epsilonhat}(\thetabar), ~ \qt \in N_{\delta}\(\EQ(\thetabar)\), ~ \forall \t\) \geq \pro\(\begin{array}{l}
|\thetat(\s)-\thetabar(\s)|<\frac{\epsilonhat}{|\S|}, ~ \forall \s \in [\thetabar],~ \forall \t \text{ and}\\
\thetat < \rhotwo,~ \forall \s \in \Sbar,~ \qt \in N_{\delta}\(\EQ(\thetabar)\), ~\forall \t
\end{array}\)\\
&= \pro\(\thetat(\s)<\rhotwo, \forall \s \in \Sbar, \forall \t\) \cdot \pro\(\left.\begin{array}{l}
|\thetat(\s)-\thetabar(\s)|<\frac{\epsilonhat}{|\S|}, \forall \s \in [\thetabar], \forall \t\\
\text{and }\qt \in N_{\delta}\(\EQ(\thetabar)\), ~\forall \t
\end{array}\right\vert
\begin{array}{l}
\thetat(\s)<\rhotwo.\\
\forall \s \in \Sbar, \forall \t\end{array}\)
\end{align*}
For any $\thetazero \in N_{\epsilon^1}(\thetabar)$ and any $\q^1 \in N_{\loadep}\(\EQ(\thetabar)\)$, we know from Lemmas \ref{lemma:stopping_time} -- \ref{lemma:other_parameters} that: 
\begin{align*}
&\pro\(\thetat(\s)<\rhotwo, ~\forall \s \in \Sbar, ~\forall \t\)>\gamma, \text{ and } \\
&\pro\(\left.\begin{array}{l}
|\thetat(\s)-\thetabar(\s)|<\frac{\epsilonhat}{|\S|}, ~ \forall \s \in [\thetabar], ~\forall \t\\
\text{and }\qt \in N_{\delta}\(\EQ(\thetabar)\), ~\forall \t
\end{array}\right\vert
\thetat(\s)<\rhotwo, ~\forall \s \in \Sbar, ~\forall \t\) =1
\end{align*}
Therefore, for any $\thetazero \in N_{\epsilon^1}(\thetabar)$ and any $\q^1 \in N_{\loadep}\(\EQ(\thetabar)\)$, the states of learning dynamics satisfy $\lim_{\t \to \infty} \pro\(\thetat\in N_{\thetaepfin}(\theta), ~ \qt \in  N_{\loadepfin}\(\EQ(\thetabar)\) \)> \gamma$. Thus, $\(\thetabar, \qbar\)$ is locally stable.
\QEDA